\documentclass{elsart}
\usepackage[dvips,dvipdf]{graphicx}
\usepackage{amssymb}

\newcommand{\cd}{\makebox[0.08cm]{$\cdot$}}

\begin{document}

\begin{frontmatter}

\title
{Many-body Fock sectors in Wick-Cutkosky model}
\author[seoul]{Dae~Sung~Hwang\thanksref{label1}}
\and
\author[moscow]{V.A.~Karmanov\corauthref{cor}\thanksref{label2}}
\address[seoul]{Department of Physics, Sejong University, Seoul 143--747,
Korea}
\address[moscow]{Lebedev Physical Institute, Leninsky Prospekt 53,
119991 Moscow, Russia}
\thanks[label1]{e-mail: dshwang@sejong.ac.kr, dshwang@slac.stanford.edu}
\thanks[label2]{e-mail: karmanov@sci.lebedev.ru, karmanov@lpsc.in2p3.fr}
\corauth[cor]{Corresponding author.}
\bibliographystyle{unsrt}

\begin{abstract}
In the model where two massive scalar particles interact by the ladder
exchanges of massless scalar particles (Wick-Cutkosky model), we study in
light-front dynamics the contributions of different Fock sectors  (with
increasing number of exchanged particles) to full normalization integral
and electromagnetic form factor. It turns out that two-body sector always
dominates. At small coupling constant $\alpha\ll 1$, its contribution is
close to 100\%. It decreases with increase of $\alpha$. For maximal value
$\alpha=2\pi$, corresponding to the zero bound state mass, two-body sector
contributes to the normalization integral 64\%, whereas the three-body
contribution is 26\% and the sum of all higher contributions from four- to
infinite-body sectors is 10\%. Contributions to the form factor from
different Fock sectors fall off faster for asymptotically large $Q^2$,
when the number of particles in the Fock sectors becomes larger. So,
asymptotic behavior of the form factor is determined by the two-body Fock
sector.
\end{abstract}

\begin{keyword}
Light-front dynamics \sep Fock sectors
\sep electromagnetic form factors

\PACS  11.10.St \sep 11.15.Tk \sep 13.40.Gp
\end{keyword}
\end{frontmatter}

\section{Introduction}
In field theory, the decomposition of the state vector in the basis of
free fields with given momenta results in the concept of relativistic wave
function in momentum space. The latter is the coefficient of the
decomposition -- the Fock component. The state vector is described by an
infinite set of the Fock components, corresponding to different numbers of
particles.

However, the models used in applications usually consider only few-body
Fock components. The infinite set of the Fock components is truncated to
one component only, with two or three quarks. The belief that a given Fock
sector dominates is mainly based on intuitive expectations and
"experimental evidences" rather than on field-theoretical analysis.
Without this analysis a model remains to be phenomenological. Though one
can perturbatively estimate the next Fock components (with one or even few
extra particles), this says nothing about convergence of full Fock
decomposition.

The difficulty to take into account the many-body sectors is
caused by the fact that it is equivalent to solving a true
field-theory problem. Important part of this problem -- finding
the Bethe-Salpeter (BS) function \cite{BS} -- is solved in the
Wick-Cutkosky model \cite{wcm}. In this case the BS equation is
reduced to the one-dimensional one which can be easily solved
numerically for arbitrary coupling constant. In weak and strong
coupling limits there are explicit analytical solutions. This
allows us to use Wick-Cutkosky model for non-perturbative study
the Fock decomposition.

In the Wick-Cutkosky model two spinless constituents with mass $m$
interact by exchange by a massless scalar particle. Corresponding BS
function is the "two-body" one in the sense that it depends on two
four-dimensional particle coordinates. The term "two-body" here is rather
slang than reflection of a real physical situation. It may be misleading.
We emphasize that this {\em does not} mean that the "two-body" BS function
describes a two-body system, i.e., a system with the state vector
truncated to the two-body Fock component. On the contrary, the
corresponding state vector contains infinite set of the Fock components
with two massive constituents and $0,1,2,\ldots,\infty$ massless exchanged
particles.
\begin{figure}[!ht]
\centering
\includegraphics{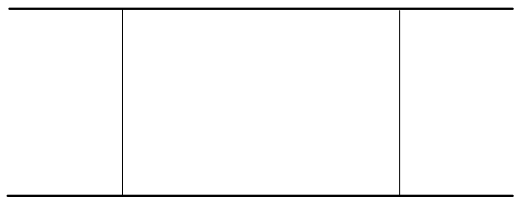}
\caption[*]{Feynman ladder graph with two exchanges.}
\label{feyn2}
\end{figure}

\begin{figure}[!ht]
\centering
 \includegraphics{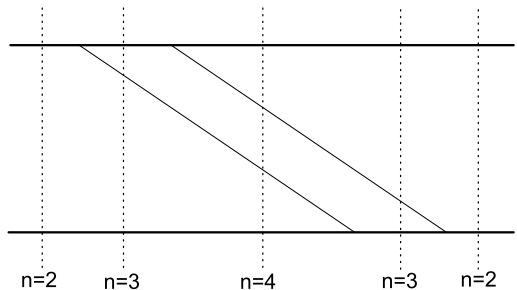}
\caption[*]{One of six time-ordered ladder graphs, generated by
the Feynman graph Fig.~\protect{\ref{feyn2}}.}
 \label{LF2}
\end{figure}
\par
This can be explained as follows. In terms of the Feynman
diagrams, the one-boson exchange kernel generates the ladder
graphs only. One of these graphs, with two exchanges, is shown in
Fig.~\ref{feyn2}. We will work in the light-front dynamics (LFD).
In terms of the time-ordered graphs, in the light-front (LF) time,
 this Feynman diagram contributes to the two-, three- and
four-body states (with two massive constituents and with zero, one
and two exchanged bosons). There are six such time-ordered graphs.
One of these graphs is shown in Fig.~\ref{LF2}. Since the BS
equation deals with infinite series of ladder exchanges, it
implicitly incorporates all the intermediate states with number of
particles from two to infinity. The problem (solved in the present
paper) is to "extract" from the BS solution the two-body
contribution and total contribution of all other Fock components.
Comparison of total contribution with the two-body one shows the
contribution of the many-body sectors.

Because of the ladder kernel, the Wick-Cutkosky model is still
approximate. All the exchange particles, emitted by one and the
same boson, must be and absorbed by other one. The graphs with
particles emitted by different bosons in the same intermediate
state are generated by the cross Feynman box which is absent in
the Wick-Cutkosky model. Self-energy graphs are absent too. In
spite of these restrictions, the Wick-Cutkosky solution does
contain in the state vector an infinite sum of many-body Fock
components. Namely this property is most important for our work.

In Wick-Cutkosky model there are also some abnormal solutions
which have no counterpart in the non-relativistic potential theory
\cite{wcm,nak69}. There was a discussion \cite{nak69}, whether
they are physical ones or not. This is a separate problem, not
related to our study. We  deal with the normal solution. In
non-relativistic limit the two-body Fock component, which only
survives in the state vector of normal solution, turns into the
usual ground state wave function in the Coulomb potential.

A truncated LF field theory was developed in \cite{phw1990}. Some
estimations of the higher Fock sectors were done in \cite{SBK}. Truncated
LF Fock space decomposition was applied in \cite{hb1999} to
nonperturbative study the large $\alpha$ QED. Recently, the Fock
decomposition, incorporating first three Fock sectors,  was used for
nonperturbative renormalization in a scalar model \cite{bckm} and, with
two sectors, -- in Yukawa model \cite{glp1992,KMS} and in gauge theory
\cite{KMS}.

In the present paper, we study in the Wick-Cutkosky model the
contribution of the many-body sectors in electromagnetic form
factor and compare it with the two-body contribution. Besides, we
calculate the contribution to the normalization integral of the
two- and three-body states. Subtracting them from 1, we get
contribution of all the states with $n\ge 4$. In this way we
investigate the convergence of the Fock decomposition.
Calculations are carried out nonperturbatively in full range of binding energy
$0\le B\le 2m$.

Plan of the paper is the following. In Section \ref{Fock} we define
the Fock decomposition (in the framework of LFD), and remind the
definition of the BS function. In Section \ref{gzM} the BS equation
is solved. In addition to analytical solutions in limiting cases
$M\to 2m$ and $M=0$, found in \cite{wcm}, we find, for small
$\alpha$, next correction of the order of $\alpha\log\alpha$ and solve BS
equation numerically for any $\alpha$. In Section \ref{two-body} the
two-body contribution to the normalization integral and to the
electromagnetic form factor is calculated for both
small and large coupling constants. The form factor asymptotic at
$Q^2\to 0$ is obtained. In Section \ref{three-body} the same is done
for three-body contribution and in Section \ref{Full} for sum of
all Fock sectors. Sect. \ref{num}  is devoted to numerical
calculations for arbitrary coupling constant. Sect. \ref{summ}
contains  summary and discussion. Some technical details are given
in Appendices \ref{correctWC}, \ref{app} and \ref{ap_ffs}.

\section{Fock decomposition and BS function}\label{Fock}

The state vector $\left\vert p \right>$, eigenstate of full
Hamiltonian, can be decomposed  in terms of the eigenstates of
free Hamiltonian. This results in natural formulation of the
concept of a relativistic wave function in terms of the LF Fock
expansion at fixed LF time, which is usually put to zero for a
stationary state:  $\omega\cd x=0$. The null four-vector $\omega$
($\omega^2=0$) determines the orientation of the LF plane; the
freedom to choose $\omega$ provides an explicitly covariant
formulation of LF quantization~\cite{cdkm}. The Fock decomposition
has the form:
\begin{eqnarray}
\left\vert p \right> &=& \sum_{n \ge 2} \int
\psi(k_1,\ldots,k_n,p,\omega\tau) \nonumber\\ &&\times
\delta^{(4)}\left(\sum^n_j k_j-p-\omega\tau\right) 2(\omega\cd p)
\d\tau \;\prod^{n}_{i=1} \frac{\d^3k_{i}}{(2\pi)^3
2\varepsilon_{k_i}} a^{\dagger}(\vec{k}_i)\left| 0 \right>.
\label{stv}
\end{eqnarray}
Here $a^\dagger$ is the usual creation operator and
$\varepsilon_{k_i}=\sqrt{m_i^2+\vec{k}_i^{\,2}}$. All the
four-momenta are on the corresponding mass shells: $k_j^2=m_j^2$,
$p^2=M^2$, $(\omega\tau)^2=0$. In QCD, the set of LF Fock state
wave functions $ \psi(k_1,\ldots,k_n,p,\omega\tau)$ represents the
ensemble of quark and gluon states possible when the hadron is
intercepted at the light-front. In Wick-Cutkosky model this set
represents the states with two massive constituents
$+0,1,2,\ldots$ massless exchange particles, forming \mbox{two-,}
\mbox{three-,}
four-body states, etc. The constituents are spinless and we
consider the state $\left\vert p \right>$ with total zero angular
momentum. Therefore we omit the spin indices. The scalar variable
$\tau$ controls the off-shell continuation of the wave function.
From the point of view of kinematics, the four-momentum
$\omega\tau$ can be considered on equal ground with the particle
four-momenta $k_1,\ldots,k_n,p$. Being expressed through them (by
squaring the equality $\sum^n_i k_i=p+\omega\tau$), $\tau$ reads:
$ \tau=(M_0^2-M^2)/(2\omega\cd p)$, where $M_0^2=(\sum^n_i
k_i)^2$.

For system with zero total angular momentum the wave functions
$\psi$ are the scalar functions and they depend on a set of scalar
products of the four-momenta $k_1,\ldots,k_n,p,\omega\tau$ with
each other. One should choose a set of $3n-4$ independent scalar
products. A convenient way to choose these variables is the
following. We define
\begin{equation}\label{sc8}
x_i=\omega\cd k_i/\omega\cd p\ , \quad R_i=k_i-x_i p
\end{equation}
and represent the spatial part of the four-vector
$R_i=(R_{i0},\vec{R}_i)$ as $\vec{R}_i=\vec{R}_{i\|}
+\vec{R}_{i\perp}$, where $\vec{R}_{i\|}$ is parallel to
$\vec{\omega}$  and $\vec{R}_{i\perp}$ is orthogonal to
$\vec{\omega}$. In the standard version of LFD the difference
$\sum_i k_i-p$ is non-zero for the minus component only. In
covariant formulation this means: $\sum_i k_i=p+\omega\tau$.
Therefore: $\sum_i \vec{R}_{i\perp}=0$, $\sum_i x_i=1$. Since
$R_i\cd\omega=R_{i0}\omega_0-\vec{R}_{i\|}\cd\vec{\omega}=0$ by
definition  of  $R_i$, it follows that $R_{i0}=|\vec{R}_{i\|}|$,
and, hence,  $\vec{R}^2_{i\perp} =-R_i^2$ is Lorentz and rotation
invariant. Similarly one can show that $\vec{R}_{i\perp} \cd
\vec{R}_{j\perp}=-R_i\cd R_j$. Therefore the scalar products
$\vec{R}_{i\perp} \cd \vec{R}_{j\perp}$ are also Lorentz and
rotation invariants. In terms of these variables the invariant
energy $s\equiv M_0^2=(\sum_i k_i)^2$ is given by
\begin{equation}\label{sint}
s=\sum_i \frac{\vec{R}^2_{i\perp}+m_i^2}{x_i}.
\end{equation}
The variables $\vec{R}_{i\perp},x_i$ are analogous to the
well-known variables in the infinite momentum frame \cite{sw}.

The wave functions are parametrizd as:
$\psi=\psi(\vec{R}_{1\perp},x_1;\vec{R}_{2\perp},x_2;
\ldots;\vec{R}_{n\perp},x_n)$ and should depend on $x_i$,
$\vec{R}^2_{i\perp}$ and on $\vec{R}_{i\perp}\cd
\vec{R}_{j\perp}$. In terms of these variables, the integral in
Eq.~(\ref{stv}) is transformed as
\begin{eqnarray*}
&&\lefteqn{\int \ldots \delta^{(4)}\left(\sum^n_j
k_j-p-\omega\tau\right)
    2(\omega\cd p) \d\tau
\;\prod^{n}_{i=1} \frac{\d^3k_{i}}{(2\pi)^{3}2\varepsilon_{k_i}}}\\
&&= \int \ldots 2\delta\left(\sum_j^n x_j-1\right)
\delta^{(2)}\left(\sum_j^n \vec{R}_{j\perp}\right) \prod^{n}_{i=1}
\frac{\d^2 R_{i\perp} \d x_i}{(2 \pi)^3 2x_i}.
\end{eqnarray*}

The state vector (\ref{stv}) enters the definition of the BS
function:
\begin{equation}\label{bsa}
{\mathit
 \Phi}(x_1,x_2,p)=\left\langle 0
\right|T(\varphi(x_1)\varphi(x_2))\left|p\right\rangle.
\end{equation}
Since the function ${\mathit \Phi}(x_1,x_2,p)$ depends on two
four-dimensional coordinates $x_1,x_2$, one often refers to the
function (\ref{bsa}) as describing  a two-body system. However,
the state vector $\left|p\right>$ is the full state vector.
 It contains all
the Fock components, with the constituent numbers
$n=2,3,4,\ldots$. The operators $\varphi(x_1),\varphi(x_2)$ in
(\ref{bsa}) are the Heisenberg operators (of the massive field).

We need the BS function $\Phi(k,p)$ in the momentum space. It is
related to (\ref{bsa}) as follows:
\begin{eqnarray}
&&{\mathit \Phi}(x_1,x_2,p)=(2\pi)^{-3/2}\exp\Bigl(-ip\cd
(x_1+x_2)/2\Bigr) \tilde{\mathit \Phi}(x,p),\nonumber\\
&&\Phi(k,p)=\int \tilde{\mathit \Phi}(x,p) \exp(ik\cd x) \d^4x,
\label{bsimp}
\end{eqnarray}
where $x=x_1-x_2$ and $p=p_1+p_2$ is the on-mass-shell momentum of
the bound state ($p^2=M^2$),  $p_1,p_2$ are the off-shell momenta
($p_1^2\neq p_2^2\neq m^2$) and $k=(p_1-p_2)/2$.

Knowing the BS function (\ref{bsimp}), we can extract from it the
two-body component \cite{cdkm}:
\begin{equation} \label{bs8}
\psi(\vec{R}_{\perp},x) =\frac{(\omega\cd k_1 )(\omega\cd k_2
)}{\pi (\omega\cd p)}\int_{-\infty }^{+\infty }\Phi
(k+\beta\omega,p)\d\beta.
\end{equation}
This relation is independent of any model.

In the Wick-Cutkosky model, the BS function $\Phi(k,p)$ for the
ground state with zero angular momentum $L=0$  has the following
integral representation \cite{wcm,nak69}:
\begin{equation}
\Phi (k,p)=-\frac{i}{\sqrt{4\pi }}\int_{-1}^{+1}
\frac{g_M(z)\d z}{(m^2-M^2/4-k^2-zp\cd k-i\epsilon)^3}\ .
\label{bs8p}
\end{equation}
This representation is valid and exact for zero-mass exchange. The
 function $g_M(z)$ is determined by an integral equation
\cite{wcm,nak69} given below in Section \ref{gzM}. It is solved
analytically in the limiting cases of asymptotically small and
extremely large binding energy and numerically -- for any binding
energy.

Substituting (\ref{bs8p}) into (\ref{bs8}), we find the two-body
Fock component \cite{karm80,Saw}:
\begin{equation}\label{bs10}
\psi(\vec{R}_{\perp},x) =\frac{x(1-x)g_M(1-2x)}{2\sqrt{\pi }
\Bigl(\vec{R}_{\perp}^2+m^2-x(1-x)M^2\Bigr)^2}\ .
\end{equation}

When we express electromagnetic form factor through the BS
function, we do not make any truncation of the Fock space. We get
sum of contributions of all the Fock components. On the other
hand, having found the two-body component (\ref{bs10}), we can
calculate its contribution to form factor and to the full
normalization integral. In addition, below we will calculate
explicitly the three-body contribution. Comparison of full
contribution with the two-body ones allows to find the many-body
sector contribution, for arbitrary values of coupling constant.
Comparison of two-body,  three-body and full contributions allows
to see, how the Fock decomposition converges. In the present paper
we will realize this program.

We will use the BS function (\ref{bs8p}) and the LF two-body wave
function (\ref{bs10}) to calculate full form factor (and
normalization integral) and two-body contribution to them. For
these calculations we need to know the function $g_M(z)$.

\section{Normalization condition for $g_M(z)$}\label{Norm}
Since the state vector $|p\rangle$ in definition (\ref{bsa}) of
the BS function is normalized, the BS function does not contain
any arbitrary factors and, hence, is also properly normalized. The
normalization factor cannot be found, of course, from the
homogeneous BS equation, but is determined by a special
normalization condition \cite{nak69}. It is equivalent to the
condition based on the charge conservation \cite{nak69}, which
means that $F_{full}(0)=1$. This condition determines the
normalization of $g_M(z)$.
\begin{figure}[!ht]
\centering
 \includegraphics{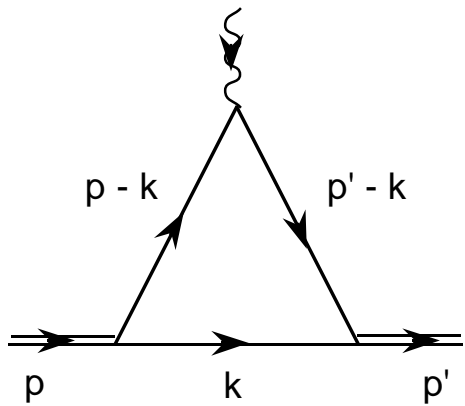}
\caption[*]{Feynman diagram for the EM form factor.}
 \label{triangle}
\end{figure}

The electromagnetic form factor $F_{full}(Q^2)$ ($Q^2=-q^2\ge 0$
and $q=p'-p$) is shown graphically in Fig.~\ref{triangle}. The
vertices at the left- and right-hand sides, being multiplied by
the propagators, are the BS functions (\ref{bsimp}). Therefore the
electromagnetic vertex is expressed in terms of the BS function as
\cite{cdkm}
\begin{eqnarray}\label{ffbs}
&&(p+p')^\mu F_{full}(Q^2) \\ &=&-i\int {\d^4k\over (2\pi)^4}\
(p+p'-2k)^\mu \; (m^2-k^2)\; \Phi \left(\frac{1}{2}p
-k,p\right)\Phi  \left(\frac{1}{2}p'-k,p'\right). \nonumber
\end{eqnarray}
 We multiply both
sides of  (\ref{ffbs}) by $(p+p')_{\mu}$, insert (\ref{bs8p}) into
(\ref{ffbs}), use the Feynman parametrization: $$
\frac{1}{a^3b^3}=\int_0^1\frac{30u^2(1-u)^2\d u} {\Bigl(a
u+b(1-u)\Bigr)^6}, $$ integrate (using Wick rotation) over
four-vector $k$ and replace the variables $z=2x-1$, $z'=2x'-1$. In
this way, at $Q=0$ we find the normalization condition:
\begin{eqnarray}\label{Nfull}
F_{full}(0)&=&-\frac{1}{2^5\pi^3}\; \int_{0}^1 g_M(2x-1)\d x\;
\int_{0}^1 g_M(2x'-1)\d x'\\ &\times&\int_0^1 {u}^2\ (1-u )^2\d u
\;\frac{\Bigl((6\xi-5)m^2+2\xi(1-\xi)M^2\Bigr)}
{\Bigl(m^2-\xi(1-\xi)M^2\Bigr)^4}=1,
 \nonumber
\end{eqnarray}
where $ \xi=xu+x'(1-u).$

\section{Solution of Wick-Cutkosky equation}\label{gzM}
The integral equation for the function $g_M(z)$ for the ground
state reads \cite{wcm,nak69}:
\begin{equation}\label{g0}
g_M(z)=\frac{\alpha}{2\pi}\int_{-1}^1 K(z,z')g_M(z')\d z'
\end{equation}
with the kernel:
\begin{equation}\label{K}
K(z,z')= \frac{m^2}{m^2-\frac{1}{4}(1-{z'}^2)M^2}\left[
\frac{(1-z)}{(1-z')}\theta(z-z')+\frac{(1+z)}{(1+z')}\theta(z'-z)\right]
.
\end{equation}
Here $\alpha=g^2/(16\pi m^2)$ and $g$ is the coupling constant in
the interaction Hamiltonian $H^{int}=-g\varphi^2(x)\chi(x)$. In
nonrelativistic limit the interaction is reduced to the Coulomb
potential $V(r)=-\frac{\alpha}{r}$. The function $g_M(z)$ is
defined in the interval $-1\leq z \leq 1 $ and it is even:
$g_M(-z)=g_M(z)$.

At small binding energy $B=2m-M$, and, on the contrary, at
extremely large binding energy equal to $2m$, the solutions
$g_M(z)$ are found explicitly.
At asymptotically small binding energy (corresponding to
$\alpha\to 0$) it has the form \cite{wcm}:
\begin{equation}\label{bs11}
g_{M\to 2m}(z)=N_{M\to 2m}(1-|z|)
\end{equation}
and the binding energy is the non-relativistic one in the Coulomb
potential: $B=\frac{m\alpha^2}{4}$. The normalization constant
$N_{M\to 2m}$ is found by substituting $g_{M\to 2m}(z)$ into the
normaization condition (\ref{Nfull}). Calculation gives:
\begin{equation}\label{N2m}
N_{M\to 2m}= 8\sqrt{2}\pi\alpha^{5/2} m^3.
\end{equation}

The higher Fock sectors are generated by extra exchanges which
contain extra degrees of $\alpha$. They are omitted in the
solution (\ref{bs11}). Therefore, to analyze the many-body
contributions, we should take into account next $\sim \alpha$
correction to Eq.~(\ref{bs11}).  To find it, we use the method of
paper \cite{FFT}.  The correction contains both the terms $\sim
\alpha$ and $\sim \alpha\log\alpha$. We keep the leading  $\sim
\alpha\log\alpha$ only. The details are given in Appendix
\ref{correctWC}. The solution of Eq.~(\ref{g0}) obtains the form:
\begin{equation}\label{g1}
g_{M}(z)=N\left[1-|z|+\frac{\alpha}{2\pi}(1 +
|z|)\log(z^2+\alpha^2/4)\right] .
\end{equation}
Corresponding binding energy $B=2m-M$ is given by \cite{FFT}:
\begin{equation}\label{B}
B=\frac{m\alpha^2}{4}-\frac{m\alpha^3}{\pi}\log\frac{1}{\alpha} .
\end{equation}
The normalization factor $N$ in (\ref{g1}) (still found from the
condition (\ref{Nfull})) now reads:
\begin{equation}\label{N1}
N= 8\sqrt{2}\pi\alpha^{5/2}
m^3\left[1+\frac{5\alpha}{\pi}\log\alpha\right].
\end{equation}

In the opposite limiting case $B=2m,M=0$, which is achieved at
$\alpha=2\pi$, the solution has the form \cite{wcm,nak69}:
\begin{equation}\label{M0}
g_{M=0}(z)=N_{M=0}(1-z^2)\ .
\end{equation}
The normalization coefficient reads:
\begin{equation}\label{N0}
N_{M=0}= 6\sqrt{30}\pi^{3/2}m^3.
\end{equation}
The latter case is especially simply checked. We substitute Eq.
(\ref{M0}) with the normalization (\ref{N0}) into Eq.
(\ref{Nfull}) and find:
 $$ F_{full}(0)= 540
\int_0^1 u^2(1 - u)^2 x(1 - x) x'(1 - x') (5 - 3x - 3x')\;\d u \d x
\d x'=1. $$ This justifies the value (\ref{N0}).

 The BS functions
corresponding to the small and large binding energy solutions
(\ref{bs11}) and (\ref{M0}) are given explicitly in Appendix
\ref{app}.

For arbitrary bound state mass  $M$ we solve equation (\ref{g0})
numerically and find corresponding function $g_M(z)$ and coupling
constant $\alpha$. The interval $-1 \le z\le 1$ is split in 50
equal subintervals. In every subinterval the function $g_M(z)$ is
represented by a sum of the quadratic spline functions. We
substitute a given initial $g_M(z)$ in the right-hand side of
(\ref{g0}), represent  the result of numerical integration again
through the spline functions and iterate. For $M\to 2m$ and $M\to
0$ we numerically reproduce the analytical solutions (\ref{bs11})
and (\ref{M0}) correspondingly. The iterations converge very
quickly. For example, for $M= 0$, even if we start iterations with
the function (\ref{bs11}), corresponding to the opposite limit
$M\to 2m$, after a few iterations we reproduce the true solution
(\ref{M0}). The accuracy of seven digits (for example) in the
function $g_{M=0}(z)$ is achieved after 8 iterations.

\begin{figure}[!ht]
\centering
 \includegraphics{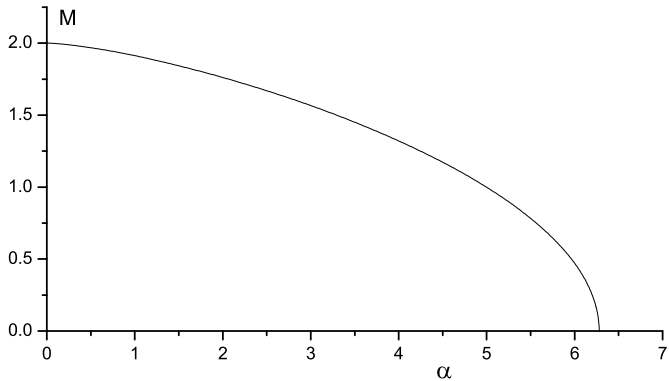}
\caption[*]{Bound state mass $M$ vs. coupling constant $\alpha$.
In all the figures we put $m=1$.} \label{M_alpha}
\end{figure}

In Fig.~\ref{M_alpha} the dependence of $M$ on $\alpha$ is shown.
In all the figures we use the units  of $m$, i.e. we put $m=1$.
One can see that $M$ turns into zero at $\alpha\approx 6.28$, that
agrees with the analytical solution $\alpha(M=0)=2\pi$.

\begin{figure}[!ht]
\centering
\includegraphics{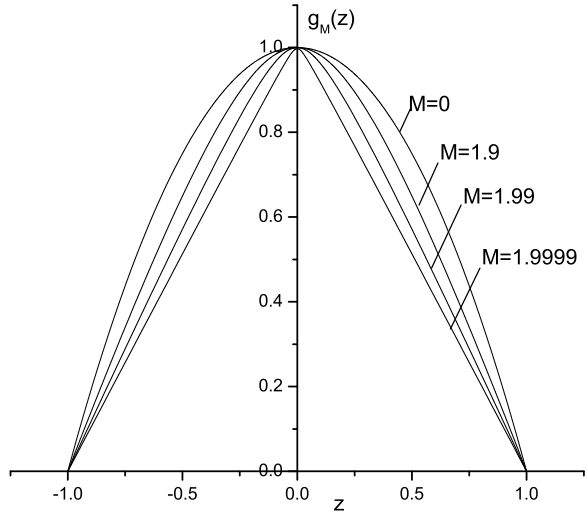}
\caption[*]{Functions $g_M(z)$ normalized by $g_M(0)=1$.}
\label{g01}
\end{figure}
\par
To compare the form of the functions $g_M(z)$ vs. $z$ for
different $M$'s, we, at first, normalize all $g_M(z)$'s  by the
condition $g_M(0)=1$ and show them in Fig.~\ref{g01} for the
values $M=0,\;1.9,\;1.99,\;1.9999$. It is seen that the form of
$g_M(z)$ does not depend significantly on $M$. For very different
values $M=0$ and $M=1.9$ the curves of $g_M(z)$ are rather close
to each other. All the curves for $0 < M < 1.9$ are between these
two curves. We emphasize that with the function $g_M(z)$
normalized by $g_M(0)=1$ the form factor $F_{full}(0)$ is not
normalized to 1.

\begin{figure}[!ht]
\centering
 \includegraphics{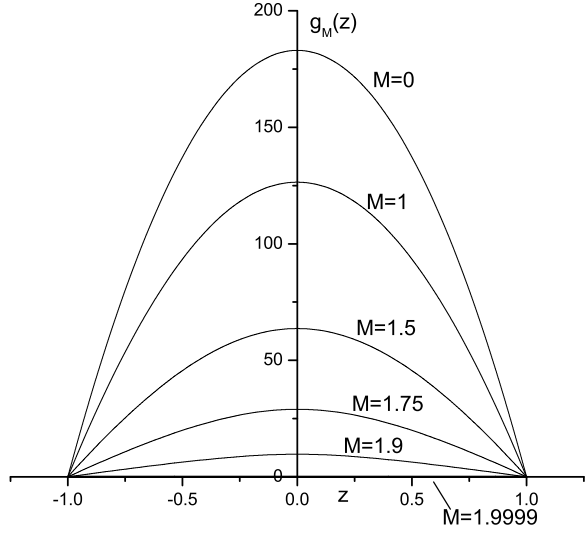}
\caption[*]{Functions $g_M(z)$ normalized by $F_{full}(0)=1$.}
\label{gF1}
\end{figure}

The functions $g_M(z)$ normalized by the condition
$F_{full}(0)=1$, Eq.~(\ref{Nfull}), are shown in Fig.~\ref{gF1}
for the values $M=0, \;1, \;1.5, \;1.75, \;1.9, \;1.9999$. They
strongly depend on $M$. This dependence comes mainly from the
normalization factor. In the scale of Fig.~\ref{gF1} the function
$g_M(z)$ at $M=1.9999$ is indistinguishable from the $z$-axis.

Now we are in position to calculate form factors and two- and
three-body contributions to the normalization integral.

\section{Two-body contribution}\label{two-body}
\subsection{Form factor}\label{2bff}
\begin{figure}[!ht]
\centering
\includegraphics{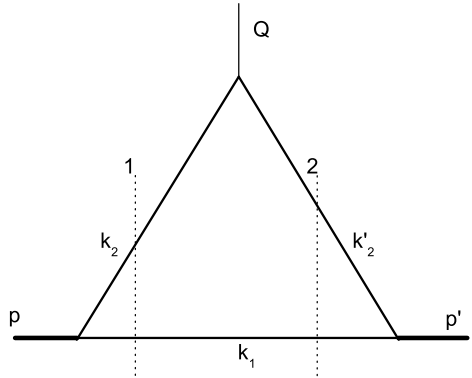}
\caption[*]{LF diagram for the two-body contribution to the EM
form factor.}
 \label{graph_2b}
\end{figure}
The two-body contribution $F_{2b}(Q^2)$ to form factor
$F_{full}(Q^2)$ is shown graphically in Fig.~\ref{graph_2b}. This
diagram corresponds to the time-ordered (in the LF time) graph
technique. The latter exists in a few versions \cite{sw,cdkm,bpp},
giving equivalent results. Dealing with the scalar particles, we
will use the Weinberg rules \cite{sw} (given in Appendix
\ref{ap_ffs}). Though the expression for two-body form factor in
terms of an overlap of the LF wave functions is well-known
\cite{cdkm,bpp,LB80,bh,bhms,BHHK}, we
derive it in Appendix \ref{ap_ffs}. Form factor reads:
\begin{equation}\label{f2b} F_{2b}(Q^2)= \frac{1}{(2\pi)^3} \int
\psi(\vec{R}_{1\perp},x_1)
\psi(\vec{R}_{1\perp}-x_1\vec{Q}_{\perp},x_1)
\frac{\d^2R_{1\perp}\d x_1}{2x_1(1-x_1)}, \end{equation} where ${\vec
Q}_{\perp}^2=Q^2$. Substituting in (\ref{f2b}) the wave function
$\psi(\vec{R}_{\perp},x)$ determined by Eq.~(\ref{bs10}) and
combining two $\vec{R}_{1\perp}$-dependent denominators by means
of the formula $$
\frac{1}{a^2b^2}=\int_0^1\frac{6u(1-u)\d u}{\Bigl(au+b(1-u)\Bigr)^4},
$$ we can easily integrate over $\vec{R}_{1\perp}$:
\begin{equation}
F_{2b}(Q^2)={1\over 2^5 \, \pi^3}\int_0^1 \int_0^1 {x(1-x)\;
g_M^2(2x-1) \d x\; u(1-u)\d u\over \Bigl(u(1-u)x^2 Q^2+
m^2-x(1-x)M^2{\Bigr)}^3} \ . \label{F2b}
\end{equation}
Calculating then the integral over $u$, we finally obtain:
\begin{eqnarray}
F_{2b}(Q^2)&=&{1\over 2^5 \pi^3 Q^6}\int_0^1
\left[4\gamma(1+\gamma)\log\frac{\sqrt{1+4\gamma}+1}{\sqrt{1+4\gamma}-1}
+(1-2\gamma)\sqrt{1+4\gamma}\right] \nonumber\\ &&
\phantom{{1\over 2^5 \pi^3 Q^6}\int_0^1} \times\frac{(1-x)
g_M^2(2x-1)\d x}{x^5\gamma(1+4\gamma)^{5/2}} ,
 \label{F2bb}
\end{eqnarray}
where $$ \gamma=\frac{m^2-x(1-x)M^2}{x^2Q^2}.$$

\subsection{Normalization integral}
\subsubsection{Small binding energy ($M\to 2m$)}
The two-body contribution to the normalization integral is found
from (\ref{F2b}) or (\ref{F2b}):
\begin{equation}\label{norm2}
N_2=F_{2b}(0)={1\over 192 \, \pi^3}\int_0^1 {x(1-x)\; g_M^2(2x-1)
\d x\over \Bigl(m^2-x(1-x)M^2{\Bigr)}^3} \ .
\end{equation}
In the limit of asymptotically small binding energy ($\alpha\to
0$) we should substitute in (\ref{norm2}) the function $g_{M\to
2m}(z)$, Eq.~(\ref{bs11}), and keep the leading term only (which
now is $\sim\alpha^0=1$). In this case we obtain a trivial result:
$$
 N_2=1.
$$
We see  that the normalization integral is saturated by the
two-body Fock sector. The higher Fock sectors are generated by
graphs with extra exchanges, each of them containing extra degree
of coupling constant $\alpha$. Their contributions are out of the
leading $\alpha$-term, which was only kept in the function
$g_{M\to 2m}(z)$. They are neglected in the limit $\alpha\to 0$,
that results in $N_2=1$.

Now we calculate $N_2$ in next to the leading order, taking into
account the correction $\sim\alpha\log\alpha$. For this aim, we
substitute in the integral (\ref{norm2}) the solution (\ref{g1}).
Calculating the integral (\ref{norm2}) and keeping the terms
$\sim\alpha\log\alpha$, we obtain:
\begin{equation}\label{N2anal}
N_{2}=1-\frac{2\alpha}{\pi}\log\frac{1}{\alpha}.
\end{equation}
Hence, for the contribution of the higher
Fock sectors with $n\ge 3$: $N_{n\ge 3}=1-N_2$, we find:
\begin{equation}\label{Ng2}
N_{n\ge 3}=\frac{2\alpha}{\pi}\log\frac{1}{\alpha}.
\end{equation}
Below, in Section \ref{NI}, we will compare Eq.~(\ref{Ng2}) for
$N_{n\ge 3}=N_3+N_4+N_5+\ldots$ with the three-body contribution
$N_3$.

\subsubsection{Extremely large binding energy ($M\to 0$)}
In this case we should substitute in (\ref{norm2}) the function
$g_{M=0}(z)$, Eq.~(\ref{M0}), and put $M=0$. This gives:
\begin{equation}\label{N2}
N_2=F_{2b}(0)=90\int_0^1 x^3(1-x)^3 \d x =\frac{9}{14}\approx 64\%.
\end{equation}
The value of many-body contribution $N_{n\ge 3}=1-N_2$ at $M=0$
($B=2m$):
\begin{equation}\label{Ng2p}
N_{n\ge 3}=1-N_2=\sum_{n=3}^{\infty} N_n=5/14 \approx 36\%
\end{equation}
is the maximal value in the model.  For $M=0$ the sum $\sum_{n=3}^{\infty}
N_n$  over the many-body Fock sectors contains increasing degrees of huge
coupling constant $\alpha=2\pi$. However it still converges (to the value
$5/14$).

\subsection{Asymptotic at $Q^2\to\infty$}
\subsubsection{Small binding energy ($M\to
2m$)}\label{small}

In the limit $M\to 2m$, form factor is obtained by substituting in
(\ref{F2bb}) the solution (\ref{bs11}) for $g_{M\to 2m}(z)$. It
was calculated, by a different method, in \cite{karm92}. The
asymptotic formula reads:
\begin{equation}\label{fas2b}
F_{2b}^{asymp}(Q^2)\approx\frac{16\alpha^4m^4}{Q^4}\left[1+
\frac{\alpha}{2\pi}\log\left(\frac{Q^2}{m^2}\right)\right].
\end{equation}
The limit $M\to 2m$ corresponds, for massless exchange kernel, to
$\alpha\to 0$. As explained above, in the leading $\alpha$ order
the contribution of higher Fock sectors is neglected. Therefore in
this approximation the two-body contribution coincides with the
full one. They coincide not only in asymptotic, but everywhere.

\subsubsection{Extremely large binding energy ($M\to
0$)}\label{large}
 Substituting in (\ref{F2b}) the expression
(\ref{M0}) for $g_{M=0}(z)$ and making replacement of variable
$u(1-u)=v$, we find at $M=0$:
\begin{equation}\label{F2bp}
 F_{2b}(Q^2)=1080m^6\int_0^1 x^3(1-x)^3\d x \int_0^{1/4}
\frac{v\d v}{(m^2+Q^2vx^2)^3\sqrt{1-4v}} .
\end{equation}
Form factor (\ref{F2bp}) is calculated explicitly:
\begin{eqnarray}\label{F2ba}
F_{2b}(Q^2)&=&\frac{180m^4}{(Q^2)^{3}} \left[60m^2-7Q^2+36m^2
\log^2\frac{\sqrt{4m^2+Q^2}+\sqrt{Q^2}}{2m}\right.\nonumber\\
&&\left.+3(Q^2-20m^2)\sqrt{1+\frac{4m^2}{Q^2}}
\log\frac{\sqrt{4m^2+Q^2}+\sqrt{Q^2}}{2m}\right].
\end{eqnarray}
At $Q\to \infty$ we find:
\begin{equation}\label{FLF}
F^{asymp}_{2b}(Q^2)=\frac{270m^4}{Q^4}
\left[\log\left(\frac{Q^2}{m^2}\right)-\frac{14}{3}\right].
\end{equation}
It is instructive to compare $F^{asymp}_{2b}(Q^2)$ with full form
factor and with the three-body one. This will be done below.

\section{Three-body contributions}\label{three-body}
\subsection{Form factor}
\begin{figure}[!ht]
\centering
 \includegraphics{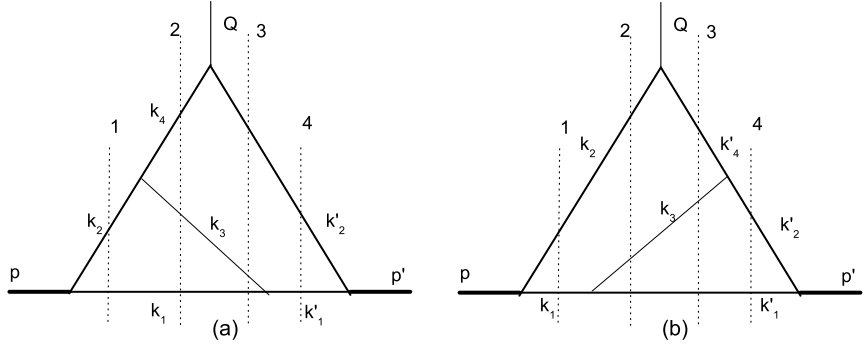}
\caption[*]{Three-body contributions to form factor.}
 \label{3b}
\end{figure}

 Comparison of two-body contribution $F_{2b}(Q^2)$ with full form factor
$F_{full}(Q^2)$ (made below analytically for asymptotic domain and
numerically for any $Q^2$) shows the total contribution to form
factor of the higher Fock sectors with $n=3,4,5,\ldots$. In order
to see the convergence of the Fock decomposition, we calculate in
this section the first term of this sum, i.e.,  the three-body
contributions. The diagrams are shown in figs. \ref{3b}~(a) and
(b).
\par
\begin{figure}[!ht]
\centering
 \includegraphics{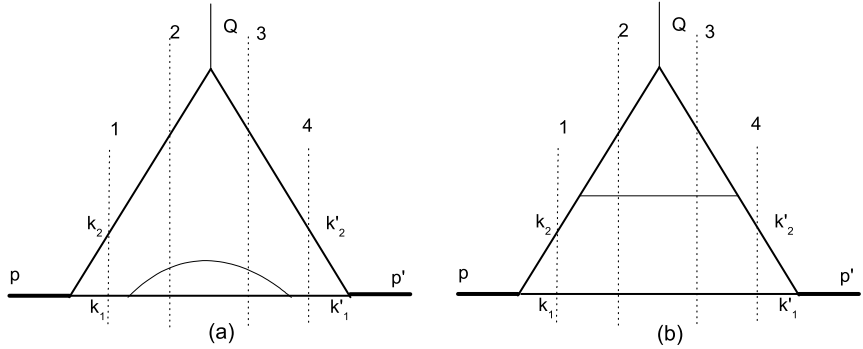}
\caption[*]{Three-body states which do not contribute to form
factor in the ladder model.}
 \label{3bp}
\end{figure}

Figures \ref{3bp}~(a) and (b) also represent the three-body
intermediate states. The graph Fig.~\ref{3bp}~(a) incorporates
self-energy, whereas the graph Fig.~\ref{3bp}~(b) can be
interpreted as a correction to the constituent form factor.
However, in the ladder approximation, they do not contribute and
therefore are omitted. There are no other three-body states.

The three-body form factor is calculated in Appendix \ref{3bf}.
The diagrams (a) and (b) in Fig.~\ref{3b} give equal
contributions. Therefore we take expression for diagram Fig.
\ref{3b}~(a) and multiply it by 2. The result has the form:
 \begin{eqnarray}\label{ffa} F_{3b}(Q^2)=\frac{\alpha
m^2}{2\pi^5}\int
\frac{\psi(\vec{R}_{1\perp},x_1)\;\psi(\vec{R'}_{1\perp},x'_1)}
{(s^{(a)}_2-M^2)\;(s^{(a)}_3-M^2)}\;\frac{\theta(x'_1-x_1)}{(x'_1-x_1)}
\nonumber\\
\times
\frac{\d^2R_{1\perp}\d x_1}{2x_1(1-x_1)}
\;\frac{\d^2R'_{1\perp}\d x'_1}{2x'_1(1-x'_1)}.
 \end{eqnarray}
The variables $s^{(a)}_2$, $s^{(a)}_3$ are the energies squared in
the intermediate states 2 and 3 in Fig.~\ref{3b}~(a). They are
expressed in terms of the integration variables $\vec{R}_{\perp}$,
$\vec{R'}_{\perp}$ and also $\vec{Q}_{\perp}$ (see Appendix
\ref{3bf}):
\begin{eqnarray}
&&s^{(a)}_2= \frac{\vec{R}_{1\perp}^2+m^2}{x_1}+
\frac{(\vec{R'}_{1\perp}-\vec{R}_{1\perp}
+x'_1\vec{Q}_{\perp})^2}{x'_1-x_1}+
\frac{(\vec{R'}_{1\perp}+x'_1\vec{Q}_{\perp})^2+m^2}{1-x'_1},
 \label{s2a}\\
&&s^{(a)}_3=
\frac{(\vec{R}_{1\perp}-x_1\vec{Q}_{\perp})^2+m^2}{x_1}+
\frac{(\vec{R'}_{1\perp}-\vec{R}_{1\perp}
+x_1\vec{Q}_{\perp})^2}{x'_1-x_1}+
\frac{\vec{R'}_{1\perp}^2+m^2}{1-x'_1} \label{s3a}.
\end{eqnarray}
We emphasize that in contrast to the two-body form factor
(\ref{f2b}), the variable $\vec{R'}_{1\perp}$ in (\ref{ffa}) is
not expressed through $\vec{R}_{1\perp}$ and $\vec{Q}_{\perp}$.
Both $\vec{R}_{\perp}$ and $\vec{R'}_{\perp}$ are independent
integration variables.

\subsection{Normalization integral}\label{NI}
Three-body contribution $N_3$ to full normalization integral is
simply $N_3=F_{3b}(0)$, that with Eq.~(\ref{ffa}) gives:
 \begin{eqnarray}\label{N3} N_3=\frac{\alpha m^2}{2\pi^5}\int
\frac{\psi(\vec{R}_{1\perp},x_1)\;\psi(\vec{R'}_{1\perp},x'_1)}
{(s_a-M^2)^2}&&\frac{\theta(x'_1-x_1)}{(x'_1-x_1)}
\nonumber\\
&&\times\frac{\d^2R_{1\perp}\d x_1}{2x_1(1-x_1)}
\;\frac{\d^2R'_{1\perp}\d x'_1}{2x'_1(1-x'_1)},
 \end{eqnarray}
 where
\begin{equation}\label{sa}
s_a= \frac{\vec{R}_{1\perp}^2+m^2}{x_1}+
\frac{(\vec{R'}_{1\perp}-\vec{R}_{1\perp})^2}{x'_1-x_1}+
\frac{\vec{R'}_{1\perp}^2+m^2}{1-x'_1} .
\end{equation}

In the leading  $\alpha\log\alpha$ order $N_3$ is calculated
analytically in Appendix \ref{norm3}. The result is the following:
\begin{equation}\label{N3s}
N_{3}=\frac{2\alpha}{\pi}\log\frac{1}{\alpha},
\end{equation}
i.e., it  coincides with $N_{n\ge 3}$, Eq.~(\ref{Ng2}). It is natural,
since the difference between $N_{n\ge 3}$ and $N_{3}$ is due to the higher
Fock sectors $N_{n\ge 4}$, which contain extra degrees of $\alpha$ omitted
in the present calculation.

The numerical calculations of $N_3$ vs. $M$ will be given in
Section \ref{num}. At $M=0$ we get: $N_3=0.257$. This illustrates the
convergence of the Fock decomposition at $\alpha=2\pi$:
$N_2\approx 64\%$, $N_3\approx 26\%$, $N_4+N_5+\ldots\approx
10\%$.

\subsubsection{Asymptotic at $Q^2\to \infty$}
As discussed above, at asymptotically small binding energy only
two-body contribution survives. Therefore we consider here the
case of extremely large binding energy only. Since
$\psi(\vec{R}_{\perp},x)$ at large $R_{\perp}$ decreases like
$\sim 1/R_{\perp}^4$ (see Eq.~(\ref{bs10})), the integral
(\ref{ffa}) well converges. Therefore we can take the limit
$Q^2\to\infty$ directly in the integrand. We put $\alpha=2\pi$,
corresponding to $M=0$. For the $Q$-dependent factor in
(\ref{ffa}) we find: $$\lim_{Q^2\to\infty}
\frac{1}{(s^{(a)}_2-M^2)(s^{(a)}_3-M^2)}=
\frac{(1-x')(x'-x)^2}{x(1-x){x'}^2}\;\frac{1}{Q^4}.$$ Then the
integral:
\begin{eqnarray*}
F^{asymp}_{3b}(Q^2)=\frac{1}{Q^4}\;\frac{m^2}{4\pi^4}
\int_0^1\d x'&&\int_0^{x'} \d x\frac{(x'-x)}{x^2(1-x)^2{x'}^4}
\\
&&\times\psi_{M=0}(\vec{R'}_{\perp},x')\d^2R'_{\perp}
\psi_{M=0}(\vec{R}_{\perp},x)\d^2R_{\perp}
\end{eqnarray*}
with the wave function $\psi_{M=0}$ given by Eq.~(\ref{psi0}) is easily
calculated and we get:
\begin{equation}\label{fas3b}
F^{asymp}_{3b}(Q^2)=\frac{180m^4}{Q^4}.
\end{equation}

\section{Full contribution}\label{Full}
\subsection{Form factor}
The contribution of all Fock components is incorporated by the
form factor $F_{full}(Q^2)$ in Eq.~(\ref{ffbs}). Its calculation
does not differ from the normalization integral explained in Section
\ref{Norm}. The results is:
\begin{eqnarray}\label{Ffull}
F_{full}(Q^2)&=&-\frac{1}{2^5\pi^3}\; \int_{0}^1 g_M(2x-1)\d x\
\int_{0}^1 g_M(2x'-1)\d x'\int_0^1 {u}^2\ (1-u )^2
\nonumber\\
&&\times
\frac{\Bigl(xx'u(1-u)Q^2+(6\xi-5)m^2+2\xi(1-\xi)M^2\Bigr)}
{\Bigl(xx'u(1-u)Q^2+m^2-\xi(1-\xi)M^2\Bigr)^4}\d u,
\end{eqnarray}
where $ \xi=xu+x'(1-u).$ At $Q=0$ form factor (\ref{Ffull}) turns
into the normalization integral (\ref{Nfull}).

\subsection{Asymptotic at $Q^2\to \infty$}\label{ffas}
As explained in Section \ref{small}, in the limit of asymptotically
small binding energy the higher Fock sector contribution
disappears. Therefore the  full form factor $F_{full}(Q^2)$
coincides with the two-body one $F_{2b}(Q^2)$, which has
asymptotic given by Eq.~(\ref{fas2b}).

We calculate asymptotic of $F_{full}(Q^2)$ in the opposite case of
extremely large binding energy. For this aim we substitute in
(\ref{Ffull}) the function $g_{M=0}(z)$ from (\ref{M0}) and make
the replacement $v=u(1-u)$. This gives:
\begin{eqnarray}\label{F2}
F_{full}(Q^2)=540m^6\int_0^{1/4} \frac{v^2
\d v}{\sqrt{\frac{1}{4}-v}}&& \int_{0}^1x(1 - x) \d x \int_{0}^1 x'(1 -
x') \d x' \nonumber\\ && \times\frac{\Bigl(m^2(5- 3x - 3x')-Q^2 x x'
v\Bigr)}{(m^2 + Q^2 v x x')^4}.
\end{eqnarray}
The integrals over $x$ and $x'$ are calculated
analytically:\footnote{$\mbox{Li}_2(z)\equiv \mbox{Polylog}(2,z)$
is a special function dilogarithm: $$
\mbox{Li}_2(z)=\int_z^0\log(1-t)\frac{dt}{t}. $$ The leading terms
of $\mbox{Li}_2(-z)$ at $z\to \infty$ are $\mbox{Li}_2(-z)\approx
-\frac{1}{2}\log^2(z+1)-\frac{\pi^2}{6}$.}
\begin{eqnarray*}
&&F_{full}(Q^2)=-\frac{540m^4}{Q^8}\int_0^{1/4}\frac{\d v}{v^2\sqrt{1-4v}}
\left[2Q^2v(3m^2+Q^2v)\right.\\ &&
\left.-(m^2+Q^2v)(6m^2+Q^2v)\log\left(1+\frac{Q^2v}{m^2}\right)-2m^2Q^2v
\mbox{Li}_2\left(-\frac{Q^2 v}{m^2}\right)\right].
\end{eqnarray*}
Now we take the limit $Q^2\to \infty$ and therefore keep in the
integrand the leading term $\propto Q^4$. We obtain:
\begin{eqnarray*}
F_{full}(Q^2)&\approx&-\frac{540m^4}{Q^4}\int_0^{1/4}\frac{\d v}{\sqrt{1-4v}}
\left[2-\log\left(1+\frac{Q^2v}{m^2}\right)\right].
\end{eqnarray*}
It is also calculated analytically.
Taking again the leading term at $Q^2\to \infty$, we find the
asymptotic:
\begin{equation}\label{FBS}
F^{asymp}_{full}(Q^2)\approx\frac{270m^4}{Q^4}
\left[\log\left(\frac{Q^2}{m^2}\right)-4\right].
\end{equation}

Comparing (\ref{FBS}) with Eq.~(\ref{FLF}) for $F_{2b}$, we see
that at $Q\to\infty$ the leading terms
$\frac{270m^4}{Q^4}\log\left(\frac{Q^2}{m^2}\right)$  of
$F_{full}$  and $F_{2b}$ , exactly coincide with each other. This
means that asymptotic of $F_{full}$ is dominated by $F_{2b}$ and
the many-body contributions decrease faster with increase of $Q^2$
than the two-body one.

Comparing Eqs.~(\ref{FLF}), (\ref{fas3b}) and (\ref{FBS}), we find
the relation
\begin{equation}\label{relat}
F^{asymp}_{full}(Q^2)=F^{asymp}_{2b}(Q^2)+F^{asymp}_{3b}(Q^2)
\end{equation}
valid for the terms $\sim 1/Q^4$. We see that the term $\sim
\frac{1}{Q^4}\log\frac{Q^2}{m^2}$ in $F^{asymp}_{full}$ comes from
$F^{asymp}_{2b}$ and the difference between $F^{asymp}_{full}$ and
$F^{asymp}_{2b}$ of the order of $\sim \frac{1}{Q^4}$ results from
$F^{asymp}_{3b}(Q^2)$. The corrections to Eq.~(\ref{relat}) result
from the higher contributions $F_{4b},F_{5b}$, etc. Since they are
absent in (\ref{relat}) in the order  $\frac{1}{Q^4}$, they
decrease more rapidly than $\frac{1}{Q^4}$. This shows a hierarchy
of asymptotic form factors: $F_{2b}>F_{3b}>F_{4b}>\ldots$. This
also indicates a good convergence of contributions of the Fock
components with increasing particle numbers, a least, in
asymptotic. In Section \ref{num} we will calculate numerically the
different contributions to the normalization integral (i.e., take
the opposite, $Q^2=0$ limit) and also find good convergence.

\subsection{Arbitrary binding energy}\label{arbit}
The momentum transfer $Q^2$ enters the equation (\ref{Ffull}) for form
factors $F_{full}$ in the combination $xx'Q^2$, and in the equation
(\ref{F2b}) for $F_{2b}$ as $x^2Q^2$. Therefore, the form factors at
$Q^2\to\infty$ are determined by the integration domain near $x,x'\to 0$
and, hence, by behavior of the function $g_M(z=2x-1)$ at binding point
$z=-1$. As shown in Section \ref{gzM}, at $z\to -1$ the function $g_M(z)$
linearly tends to zero: $g_M(z\to -1)\sim (z+1)$, independently of the
value of $M$. Therefore, the asymptotic $F_{full}(Q^2)\sim
F_{2b}(Q^2)\sim\frac{1}{Q^2}\log(\frac{Q^2}{m^2})$ is expected for any
$M$. This is natural, since this asymptotic is valid in both limiting
cases $M=0$ and $M\approx 2m$.

To check that, we represent $g_M(z)$ as: $g_M(z)=(z^2-1)f(z)$, where the
factor $(z^2-1)$ provides zeroes at $z=\pm 1$ and $f(z)$ is an even
function of $z$ which can be decomposed near $z=0$ as: $f(z)=c_0+c_2
z^2+c_4z^4+\ldots$. When we take the zero degree in this decomposition,
i.e. the term $c_0$ only, we come back to the case $g_M(z)\propto
(1-z^2)$, considered in the case $M=0$, and therefore for arbitrary
binding energy we obtain the asymptotic $F_{full}(Q^2)\sim
F_{2b}(Q^2)\sim\frac{1}{Q^2}\log(\frac{Q^2}{m^2})$ found for $M=0$. Other
terms of decomposition result in the factor
$z^{2m}{z'}^{2n}=(2x-1)^{2n}(2x'-1)^{2n}$ which does not change the
leading term. The coefficient at the leading term and next to the leading
terms depend on  particular function $f(z)$.

\section{Numerical calculations}\label{num}
The contributions $N_2$ and $N_3$ to the normalization integral from the
two- and three-body sectors   are given by Eqs.~(\ref{norm2}) and
(\ref{N3}) correspondingly. The difference:
$$ N_{n\ge 4}=1-N_2-N_3 $$
is contribution to full normalization integral of higher Fock
sectors (i.e., probability of the sectors with $n\ge 4$).

\begin{figure}[!ht]
\centering
\includegraphics{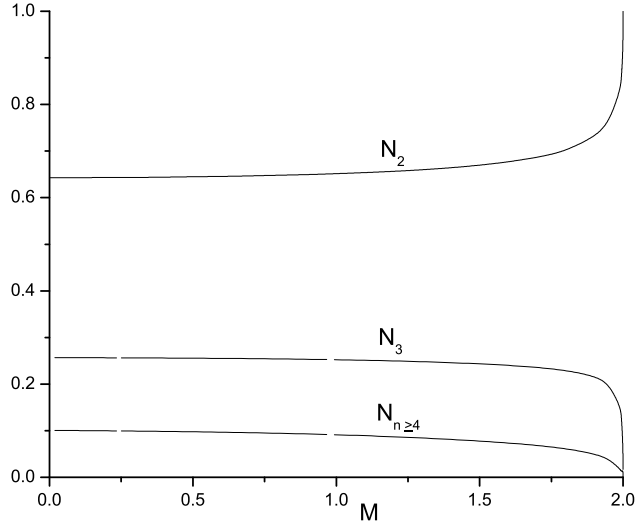}
\caption[*]{Contribution to the full normalization integral  of
the states with the constituent numbers $n=2$, $n=3$ and  $n\ge
4$.} \label{N3_M}
\end{figure}

These probabilities, i.e., $N_2$, $N_3$ and $N_{n\ge 4}$ as
function of $M$ are shown in Fig.~\ref{N3_M}. Mass $M$ (and
momentum transfer $Q$ below) are given in units of $m$, i.e., we
put $m=1$. The maximal values of $N_3$ and $N_{n\ge 4}$ are
achieved at $M=0$. In this case $N_2$ is calculated analytically
by Eq.~(\ref{N2}): $N_2=9/14\approx 0.643$.  $N_3$ is found
numerically: $N_3\approx 0.257$. Hence, the total contribution of
the Fock sectors with $n\ge 4$: $N_{n\ge
4}(M=0)=1-N_2-N_3=N_4+N_5+\ldots\approx 0.100$ is 10\% only. Like
in the case of form factor, this shows quick convergence relative
to increase of the particle number in the Fock sectors. In the
interval $0\le M \le 1.8$ these contributions are rather smooth
functions of $M$ and then, when $M$ tends to 2, $N_2$ tends to 1
and both $N_3$ and $N_{n\ge 4}$ tend to zero very quickly.

\begin{figure}[!ht]
\centering
 \includegraphics{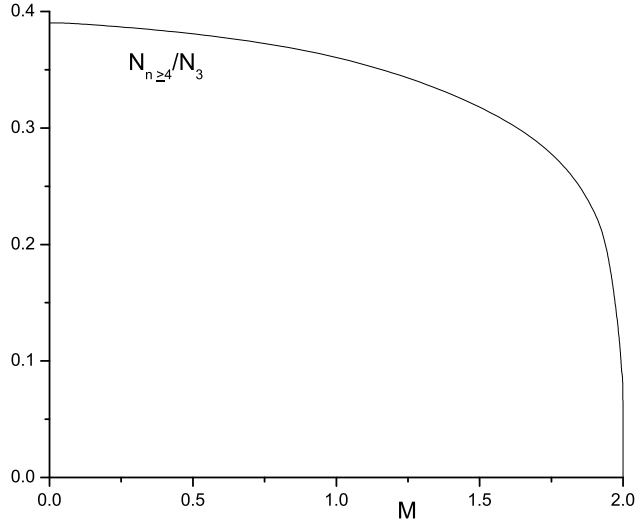}
\caption[*]{Ratio $\frac{N_{n\ge 4}}{N_3}$ vs. $M$.}
\label{N3sN2}
\end{figure}

However, $N_{n\ge 4}$ decreases faster than $N_3$. This is seen
from Fig.~\ref{N3sN2}, where the ratio $\frac{N_{n\ge 4}}{N_3}$
vs. $M$ is shown. This ratio decreases at $M\to 2$  and $N_3$
dominates over $N_{n\ge 4}$. This results from the fact that the
higher Fock sectors $N_{n\ge 4}$ contain extra degree of $\alpha$.
This is also reflected in the coincidence of analytical formulas
(\ref{Ng2}) for $N_{n\ge 3}$ with (\ref{N3s}) for $N_3$.

\begin{figure}[!ht]
\centering
 \includegraphics{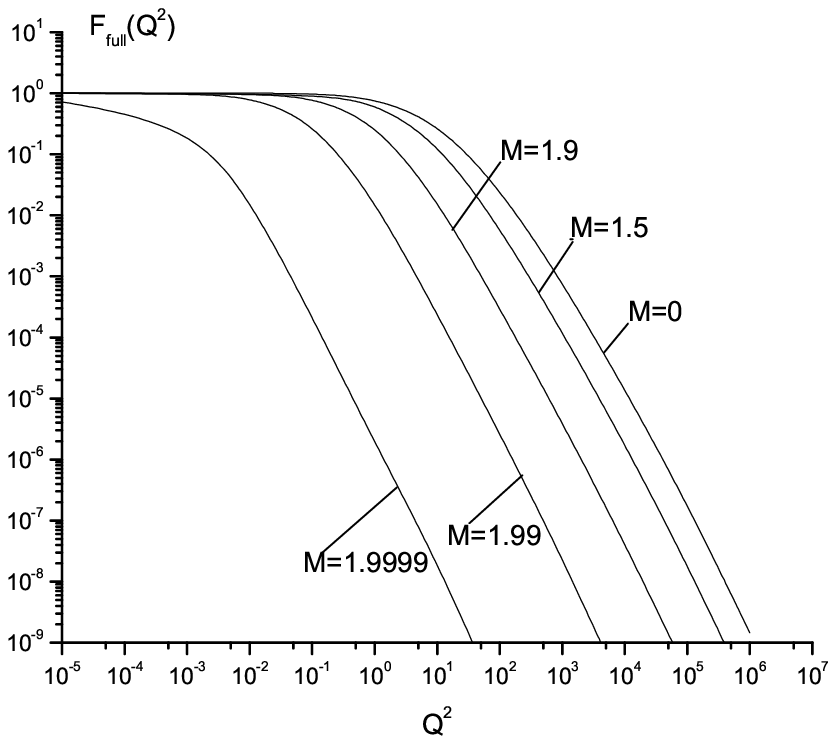}
\caption[*]{Form factor $F_{full}(Q^2)$ vs. $Q^2$ for a few
values of $M$.} \label{ffs}
\end{figure}
\par
In Fig.~\ref{ffs} the form factor $F_{full}(Q^2)$ is shown vs.
$Q^2$ for the $M$ values $M=0,$ $1.5,$ $1.9,$ $1.99,$ $1.9999$. It
is calculated by Eq.~(\ref{Ffull}).  We emphasize very strong
dependence on $M$. For example, at $Q^2=1$ the form factor for
$M=1.9999$ is many orders ($\approx 10^6$ times) smaller than the
form factor for $M=0$.
\begin{figure}[!ht]
\centering
 \includegraphics{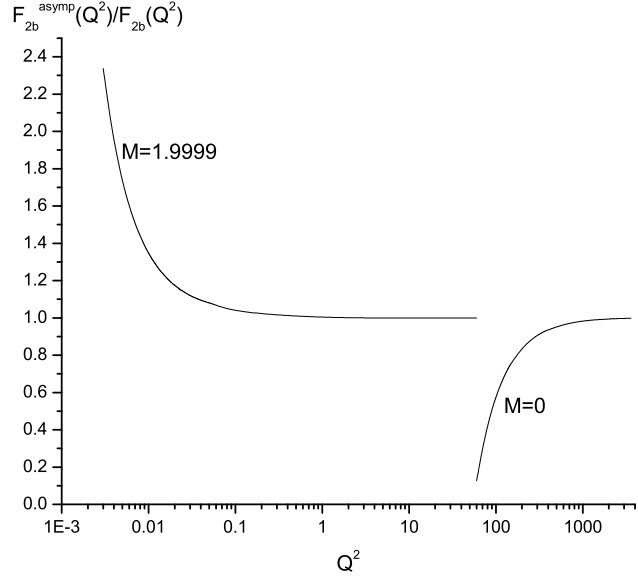}
\caption[*]{Ratio of asymptotic  of form factors
$F^{asymp}_{2b}(Q^2)/F_{2b}(Q^2)$.}
 \label{f_fas}
\end{figure}
\par
In order to show, how the two-body form factor achieves its
asymptotic value, in Fig.~\ref{f_fas} we present the ratios
 $F^{asymp}_{2b}(Q^2)/F_{2b}(Q^2)$
for $M=0$ and $M=1.9999$. For $M\to 2m$ the asymptotic form factor
$F^{asymp}_{2b}(Q^2)$ is given by Eq.~(\ref{fas2b}), and for $M=0$
-- by Eq.~(\ref{FLF}). We see that for small binding energy,
$M=1.9999$, the asymptotic is achieved much earlier (at
$Q^2\approx 0.1m^2$) than for large binding energy $M=0$
($Q^2\approx 1000m^2$).
\begin{figure}[!ht]
\centering
 \includegraphics{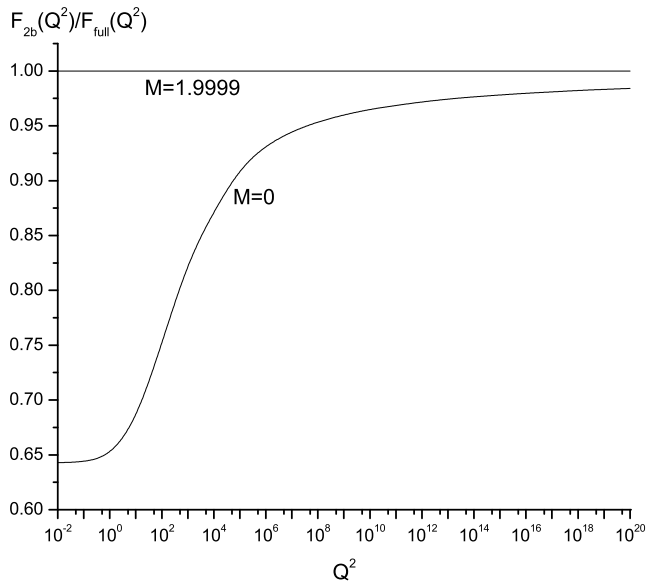}
\caption[*]{Ratio of form factors $F_{2b}(Q^2)/F_{full}(Q^2)$.}
 \label{F2bsFfull}
\end{figure}
\par
Similar situation takes place for the
ratio of full form factors $F^{asymp}_{full}(Q^2)/F_{full}(Q^2)$.
As shown in Section \ref{large}, at large values of $Q^2$ the many-body
contribution decreases more rapidly than the two-body one. Therefore at
$Q^2\to\infty$ the form factor $F_{full}(Q^2)$ coincides with
$F_{2b}(Q^2)$. The ratio of the form factors $F_{2b}(Q^2)/F_{full}(Q^2)$
vs. $Q^2$ for $M=1.9999$ and  $M=0$ is shown in Fig.~\ref{F2bsFfull}. For
very small binding energy $B=0.0001$ ($M=1.9999$), and, correspondingly,
small $\alpha$, the many body contribution is practically absent.
Therefore $F_{2b}(Q^2)/F_{full}(Q^2)\approx 1$ for any $Q^2$. For very
large binding energy $B=2$ ($M=0$) the ratio $F_{2b}(Q^2)/F_{full}(Q^2)$
tends to  1, but it becomes more or less close to 1 at huge values
$Q^2\approx 10^{20}m^2$, when $\log(Q^2/m^2)$ dominates. Both form factors
$F_{2b}(Q^2)$ and $F_{full}(Q^2)$ contain $\log(Q^2/m^2)$ with the same
coefficient but they differ by the terms without $\log(Q^2/m^2)$ (the
terms $-14/3=-4\frac{2}{3}$ in $F_{2b}(Q^2)$ in (\ref{FLF}) and $-4$ in
$F_{full}(Q^2)$ in (\ref{FBS})). The difference of form factors is
$-\frac{2}{3}$. Therefore the form factors $F_{2b}(Q^2)$ and
$F_{full}(Q^2)$ become to be close to each other very far, when
$\log(Q^2/m^2)\gg 2/3$. The relative difference
$\Bigl(F_{full}(Q^2)-F_{2b}(Q^2)\Bigr)/F_{full}(Q^2)$ decreases as $\sim
1/\log(Q^2/m^2)$ only.

\begin{figure}[!ht]
\centering
 \includegraphics{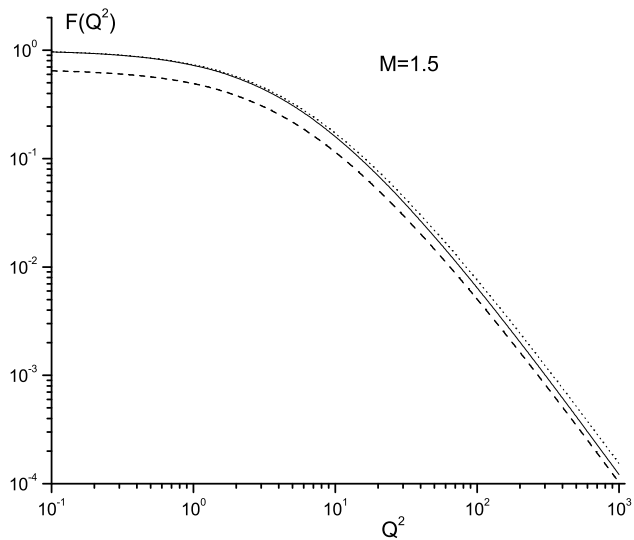}
\caption[*]{Form factors for $M=1.5$. Solid line --
$F_{full}(Q^2)$. Dash line -- $F_{2b}(Q^2)$. Dotted line --
$\tilde{F}_{2b}(Q^2)=F_{2b}(Q^2)/F_{2b}(0)$.}
 \label{FF1.5}
\end{figure}

However, in a limited domain of $Q^2$ the form factors
$F_{full}(Q^2)$ and $F_{2b}(Q^2)$ differ approximately by a
factor. It is illustrated by Fig.~\ref{FF1.5} for $M=1.5$. Solid
line in this figure represents the form factor $F_{full}(Q^2)$.
Dash line is $F_{2b}(Q^2)$. Its value at origin is:
$N_2=F_{2b}(0)=0.669$ (compare with  $N_2=F_{2b}(0)=9/14=0.643$
for $M=0$). The dotted curve shows the two-body form factor
$\tilde{F}_{2b}(Q^2)=F_{2b}(Q^2)/F_{2b}(0)$, normalized to 1 at
$Q^2=0$. It is almost indistinguishable from  $F_{full}(Q^2)$.  If
this coincidence  takes place in a more realistic model, then,
comparing the experimental data with the two-body form factor,
normalized to 1 (as it is usually done), we would conclude that
the later dominates and we would not notice 33\% many-body
contribution ($N_{n\ge 3}=1-0.669=0.331$). This would be wrong
conclusion about the structure of a system. Therefore the results
shown in Fig.~\ref{FF1.5} could be instructive.

\section{Summary and discussion}\label{summ}
The state vector in field theory does not correspond to a definite
number of particles. It is represented as a superposition of Fock
sectors with different particle numbers $n$. In Wick-Cutkosky
model this sum starts with $n=2$ and goes until infinity. We
studied the contribution to the electromagnetic form factor (and,
in particular, to the normalization integral) of the Fock sectors
with $n=2$, $n=3$ and sum of all contributions with $2\le n
<\infty$. For small and large binding energy (and coupling
constant $\alpha$) the results are obtained analytically (except
for the three-body contribution at $M=0$), and for arbitrary
binding energy - numerically.

In the limit $\alpha\to 0$ the two-body sector survives only.
Corresponding two-body Fock component is simply the
non-relativistic ground state wave function in Coulomb potential.

In the next order $\sim\alpha\log\alpha$ the higher Fock sectors contribute,
however the two-body sector still dominates. The correction $\sim
\alpha\log\alpha$ comes from the three-body contribution, the correction
$\sim \alpha^2$ results from many-body sectors with $n\ge 4$.
Since at small $\alpha$ the higher Fock sectors are suppressed by
extra degrees of $\alpha$, the Fock decomposition converges very
fast.  Not only the normalization integral, but the form factor
$F_{full}(Q^2)$, at any $Q^2$, is also dominated by the two-body
contribution.

In the leading order $\alpha\log\alpha$ the contribution of many-body
sectors to the normalization integral $N_{n\ge 3}$, extracted from the BS
function, Eq.~(\ref{Ng2}), coincides with the three-body contribution
$N_3$, Eq.~(\ref{N3s}), found by direct calculation in LFD of the
three-body intermediate state. This coincidence illustrates the main idea
of our work:  the "two-body" BS function indeed includes many-body Fock
sectors. Therefore both the two-body contribution $N_2$  and sum of
many-body contributions $N_{n\ge 3}$ can be found from a given "two-body"
BS function.

In the opposite case of large coupling constant, up to its maximal
value $\alpha=2\pi$ (corresponding to $M=0$) the Fock
decomposition still well converges. Contributions to the
normalization integral of a few Fock sectors is presented in the
Table~\ref{tab1}.
In spite of huge value of the coupling constant, contributions of
the Fock sectors quickly decrease with increase of number of
particles. Two first sectors with $n=2$ and $n=3$ determine 90\%
of the normalization integral.
\begin{table}[!ht]
\begin{center}
\begin{tabular}{|c|c|c|c|}
\hline $N_2$ & $N_3$ & $N_{n\ge 4}$ & $N_2+ N_3 +N_{n\ge 4}$\\
\hline 0.643 & 0.257 & 0.100 &1 \\ \hline
\end{tabular}
\end{center}
\caption{Contributions of the Fock sectors with the particle
numbers $n=2$, $n=3$ and $n\ge 4$  ($N_{n\ge
4}=\sum_{n=4}^{\infty}N_n$) to the full normalization integral
$N=\sum_{n=2}^{\infty}N_n=1$ of the state vector for $M=0$
($\alpha=2\pi$).}\label{tab1}
\end{table}

This can be explained as follows. Any extra Fock sector adds to the
amplitude extra factor $\sim \alpha/(s-M^2)$. There is a competition
between increasing coupling constant in numerator and increasing energy in
denominator. When $\alpha$ increases, the binding energy increases too.
Then the bound state mass $M$ decreases. This results in increase of the
energy denominators $(s-M^2)$ so that effective contribution of the factor
$\sim \alpha/(s-M^2)$ remains smaller than 1. This is a probable mechanism
of suppression of the higher Fock sectors in the Wick-Cutkosky model.

The behavior of form factors vs. $Q^2$ is also rather
instructive. The decrease of full form factor $F_{full}(Q^2)$
(Fig.~\ref{ffs}) with increase of $Q^2$ strongly depends on the
bound state mass $M$. At smaller values of $M$ the form factor
begins to decrease at larger values of $Q^2$. At $M=0$ the form
factor is almost constant up to $Q^2\approx 10m^2$ and decreases
at larger $Q^2$.

Asymptotic of form factor for the larger binding energy $B$ is
achieved at larger values of $Q^2$ (Fig.~\ref{f_fas}). For mass
$M=1.9999$ form factor $F_{2b}(Q^2)$ obtains its asymptotic form
at $Q^2\ge m^2$, whereas for $M=0$ -- at $Q^2\ge 1000m^2$.

The contribution of Fock sectors to form factor decreases with increase of
$Q^2$ more faster for higher Fock sectors (for larger values of $n$).
Therefore, in asymptotic only the two-body contribution survives (Fig.
\ref{F2bsFfull}). For small binding energy ($M=1.9999$) the two-body
component dominates for any $Q^2$. With increase of binding energy the
many-body components become more important. However, they are suppressed
at larger values of $Q^2$. The value of $Q^2$ when many-body contributions
can be neglected is larger for larger binding energy. For $M=0$ the
many-body contributions become smaller than 5\% (and the two-body form
factor constitutes 95\% of full form factor) at very large $Q^2 >
10^{10}m^2$.

The leading asymptotic terms $\sim \frac{1}{Q^4}\log\frac{Q^2}{m^2}$ in
$F_{full}$ and $F_{2b}$ are the same (with the same coefficient).
Therefore, at $Q^2\to\infty$: $F_{full}= F_{2b}$. The form factor $F_{3b}$
does not contain $ \frac{1}{Q^4}\log\frac{Q^2}{m^2}$, but it contains next
to the leading term $\sim \frac{1}{Q^4}$. With this correction $F_{full}$
at $Q^2\to\infty$ is determined by sum of two form factors, Eq.
(\ref{relat}): $F_{full}=F_{2b}+F_{3b}$, where $F_{3b}$ decreases faster
than $F_{2b}$. Contributions $F_{n\ge 4}$ of the Fock sectors with $n\ge
4$ is the next correction which decrease faster than $F_{3b}$.

Though form factor $F_{2b}$ substantially differs from $F_{full}$,
in a limited but enough large domain of $Q^2$ their ratio is a
constant. Being normalized at $Q^2=0$ to 1, $F_{2b}(Q^2)$ becomes
to be very close to $F_{full}(Q^2)$ everywhere (see Fig.~\ref{FF1.5}).
Approximate description of true form factor of a system by the
two-body one (with artificially imposed condition $F_{2b}(0)=1$)
does not mean that this system is the two-body one.

Good convergence of the Fock decomposition in Wick-Cutkosky model
indicates that the  Fock decomposition may be efficient in more
realistic field theories, which, however, should be studied
separately.

\section*{Acknowledgments}
This work is supported in part by KISTEP (D.S.H.),
and in part by the RFBR grant 02-02-16809 (V.A.K.).

\appendix
\section{Correction $\sim \alpha\log\alpha$ to the Wick-Cutkosky solution}
\label{correctWC}
 Eq.~(\ref{g0}) with the kernel (\ref{K}) is equivalent to the following
differential equation:
\begin{equation}\label{gdif}
g''_M(z)+\frac{\alpha}{\pi}\frac{m^2
g_M(z)}{(1-z^2)\left[m^2-\frac{1}{4}(1-z^2)M^2\right]}=0
\end{equation}
with the boundary conditions $g_M(\pm 1)=0$. In the limit
$\alpha\to 0$ its solution (\ref{bs11}) does not depend on
$\alpha$ (except for a normalization factor).
 To find a correction to the Wick-Cutkosky solution
(\ref{bs11}), we solve Eq.~(\ref{gdif}), using the method of the
paper \cite{FFT}. Namely, the Eq.~(\ref{gdif}) is identically
rewritten as:
\begin{equation}\label{gdif1}
g''_M(z)+\lambda (V_0+V_1)g_M(z)=0,
\end{equation}
where
 $$
\lambda=\frac{\alpha}{\eta},\quad \eta^2=\frac{4m^2}{M^2}-1,\quad
V_0=\delta(z),\quad
V_1=\frac{\eta}{\pi}\left[\frac{1+\eta^2}{(1-z^2)
(\eta^2+z^2)}\right]-\delta(z). $$ Since the lowest order solution
(\ref{bs11}) $ g^{(0)}(z)\equiv g_{M\to 2m}(z)\sim (1-|z|)$ is
obtained from (\ref{gdif1}) at $V_1=0$, i.e., it satisfies the
equation ${g''}^{(0)}(z)+\lambda^{(0)} V_0g^{(0)}(z)=0$ with
$\lambda^{(0)}=2$, $V_1$ can be considered as a perturbation.
Eq.~(\ref{gdif1}) becomes: $$ {g''}^{(0)} +{g''}^{(1)}+\ldots
+ (\lambda^{(0)}+\lambda^{(1)}+\ldots) (V_0+V_1)({g^{(0)}}
+{g^{(1)}}+\ldots)=0 ,$$ which in the first order gives \cite{FFT}:
 $$\lambda^{(1)}=
 \frac{-\lambda^{(0)}\int_{-1}^1g^{(0)}V_1g^{(0)}\d z}
 {\int_{-1}^1g^{(0)}V_0g^{(0)}\d z} $$
 and the simple equation for $g^{(1)}$:
\begin{equation}\label{gdif2} g''^{(1)}=-\left[ \lambda^{(0)}V_0 g^{(1)} +
\lambda^{(0)}V_1 g^{(0)} +\lambda^{(1)}V_0 g^{(0)}\right].
\end{equation}
From the relation $\lambda\equiv \lambda^{(0)}+
\lambda^{(1)}=\frac{\alpha}{\eta}$ with $ \lambda^{(0)}=2$ we find the
binding energy (\ref{B}) and then the solution $g^{(1)}(z)$ of Eq.
(\ref{gdif2}). It contains the terms $\sim\alpha$ and
$\sim\alpha\log\alpha$. We have found that the next order correction
$g^{(2)}(z)$ (relative to the order of the kernel $V_1$)  contains the
linear $\alpha$ contributions (i.e., the same order of $\alpha$ as
appears in $g^{(1)}(z)$). But it does not contain the terms $\sim\alpha\log\alpha$.
The binding energy calculated in \cite{FFT} by another method (in LFD)
coincides with (\ref{B}) up to the $\sim\alpha\log\alpha$ terms too.
Therefore we keep the leading $\sim\alpha\log\alpha$ term only. In this
way we find eg. (\ref{g1}) for $g_M(z)$.

\section{Explicit BS solutions and the LF wave
functions}\label{app}
 Substituting (\ref{bs11}) into (\ref{bs8p}),
we find the BS function for asymptotically small binding energy
explicitly:
\begin{eqnarray}\label{bsa1}
\Phi_{M\to 2m}(k,p)&=&\frac{-4i\sqrt{2\pi}\alpha^{5/2}m^3}
{\Bigl(m^2-\frac{1}{4}M^2-k^2-i0\Bigr)}
\\
&&\times
\frac{1}{\Bigl(m^2-(\frac{1}{2}p+k)^2-i0\Bigr)
\Bigl(m^2-(\frac{1}{2}p-k)^2-i0\Bigr)},
\end{eqnarray}
where $M=2m-\frac{m\alpha^2}{4}$.

Substituting (\ref{M0}) into (\ref{bs8p}), we find the BS function
for extremely large binding energy:
\begin{eqnarray}\label{bsa2}
\Phi_{M=0}(k,p)&=&\frac{6i\sqrt{30}\pi m^3(k^2-m^2)} {(p\cd
k)^2\Bigl(m^2-k^2-p\cd k-i0\Bigl) \Bigl(m^2-k^2+p\cd k -i0\Bigl)}
\nonumber\\ &-&\frac{3i\sqrt{30}\pi m^3}{(p\cd k)^3}
\log\frac{m^2-k^2-p\cd k-i0}{m^2-k^2+p\cd k -i0}.
\end{eqnarray}

By means of Eq.~(\ref{bs10}) we find the LF wave functions for these two
limiting cases:
\begin{equation}\label{psi2m}
\psi_{M\to 2m}(\vec{k}_{\perp},x)=\frac{8\sqrt{2\pi}\alpha^{5/2}
m^3 x(1-x)}{\Bigl(\vec{k}_{\perp}^2+m^2-x(1-x)M^2\Bigr)^2} \times
\left\{\begin{array}{ll} x,&\mbox{if $0\le x \le \frac{1}{2}$} \\
1-x,& \mbox{if $\frac{1}{2}\le x \le 1$}
\end{array}\right.
\end{equation}
\begin{equation}\label{psi0}
\psi_{M=0}(\vec{k}_{\perp},x)=\frac{12\sqrt{30}\pi m^3
x^2(1-x)^2}{\Bigl(\vec{k}_{\perp}^2+m^2\Bigr)^2} .
\end{equation}

\section{Calculating form factors}\label{ap_ffs}
\subsection{Two-body form factor}\label{2b}
To calculate the LF amplitudes, we use the Weinberg rules
\cite{sw}. To find the amplitude  $-{\mathcal M}$,
one should put in
correspondence: to every vertex -- the factor $g$ (in the theory
with interaction $H^{int}=-g\varphi^2(x)\chi(x)$), to every
intermediate state -- the factor: $ \frac{2}{s_0-s+i0}$, where $s$
is the energy squared in the intermediate state, Eq.~(\ref{sint}),
and $s_0$ is the initial (=final) state energy. In our case (bound
state): $s_0^2=M^2$. To every internal line one should put in
correspondence the factor: $ \frac{\theta(x_i)}{2x_i}$. One should
take into account the conservation laws for $\vec{R}_{i\perp}$ and
$x_i$ in any vertex and integrate over all independent variables
with the measure: $\frac{\d^2R_{i\perp}\d x_i}{(2\pi)^3}$. The
variables $\vec{R}_{i\perp},x_i$ are introduced by Eq.~(\ref{sc8})
in Section \ref{Fock}.

The LF graph determining the EM vertex of two-body state is shown
in Fig.~\ref{graph_2b}. The energies squared corresponding to the
intermediate states 1 and 2 in Fig.~\ref{graph_2b} read:
\begin{eqnarray}\label{s}
&&s_1=(k_1+k_2)^2=\frac{\vec{R}_{1\perp}^2+m^2}{x_1}
+\frac{\vec{R}_{2\perp}^2+m^2}{x_2}
=\frac{\vec{R}_{1\perp}^2+m^2}{x_1(1-x_1)},\nonumber\\
&&s_2=(k_1+k'_2)^2=\frac{\vec{R'}_{1\perp}^2+m^2}{x'_1(1-x'_1)} .
\end{eqnarray}
We take into account: $R'_1\equiv k_1-x_1 p'=R_1-x_1Q$, where
$R_1\equiv k_1-x_1 p$ and $Q=p'-p$.  Hence: $
\vec{R'}_{1\perp}=\vec{R}_{1\perp}-x_1\vec{Q}_{\perp}.$
Using the  rules \cite{sw}, we get:
\begin{eqnarray}\label{FF2}
(p+p')_{\rho}\;F_{2b}= \int (k_2+k'_2)_{\rho}&& \Gamma'\;
\frac{2}{M^2-s'}\;\frac{2}{M^2-s}\Gamma
\nonumber\\
&&\times
\frac{\theta(x_1)}{2x_1}
\frac{\theta(x_2)}{2x_2}\;\frac{\theta(x_2)}{2x_2}\;
\frac{\d^2R_{1\perp}\d x_1}{(2\pi)^3}.
\end{eqnarray}
 We used the standard condition
$\omega\cd Q=0$. Therefore $\omega\cd k_2=\omega\cd k'_2$ and,
hence, $x'_2=x_2=1-x_1$. $\Gamma$ is the vertex function related
to the wave function as:
 \begin{equation}\label{psiGam}
\psi=\frac{\Gamma}{s-M^2}.
 \end{equation}
Multiplying both sides of the equality (\ref{FF2}) by
$\omega^{\rho}$, we obtain the formula (\ref{f2b}).

\subsection{Three-body form factor and normalization integral}\label{3bf}
\subsubsection{Form factor}
We find form factor $F_{3b}$, calculating directly the amplitudes
of diagrams Fig.~\ref{3b}. This is of course equivalent to
perturbative calculation of the three-body component and then
expressing form factor through it by a standard formula.
Using the Weinberg rules \cite{sw}, similarly to two-body
contribution, Eq.~(\ref{FF2}), we obtain for the three-body graph
Fig.~\ref{3b}~(a):
 \begin{eqnarray}\label{FF3}
(p+p')_{\rho}\;F^{(a)}_{3b}&=& g^2\int (k_4+k'_2)_{\rho}\;
 \nonumber\\
&&\times
\Gamma'\; \frac{2}{(M^2-s^{(a)}_4)}\;\frac{2}{(M^2-s^{(a)}_3)}
\;\frac{2}{(M^2-s^{(a)}_2)}\; \frac{2}{(M^2-s^{(a)}_1)}\; \Gamma
 \nonumber\\
&&\times \frac{\theta(x'_1)}{2x'_1}\;
\frac{\theta(1-x'_1)}{2(1-x'_1)}\;
 \frac{\theta(1-x'_1)}{2(1-x'_1)}\;
\frac{\theta(x'_1-x_1)}{2(x'_1-x_1)}
 \nonumber\\
&&\times\frac{\theta(x_1)}{2x_1}\;
\frac{\theta(1-x_1)}{2(1-x_1)}\;
\frac{\d^2R'_{1\perp}\d x'_1}{(2\pi)^3}\;
 \frac{\d^2R_{1\perp}\d x_1}{(2\pi)^3} .
\label{c21a}
\end{eqnarray}
Here we used that $x_3=x'_1-x_1$, $x_4=x'_2=1-x'_1$, $x_2=1-x_1$.
The denominators $(M^2-s^{(a)}_1)$ and $(M^2-s^{(a)}_4)$ are
absorbed in the relation (\ref{psiGam}) between $\Gamma$ and
$\psi$. The values of $s^{(a)}_2$, $s^{(a)}_3$ are defined in
(\ref{s2a}), (\ref{s3a}).
 Multiplying both sides of Eq.~(\ref{FF3}) by
$\omega^{\rho}$, we obtain the contribution of graph Fig.
\ref{3b}~(a), which is half of  Eq.~(\ref{ffa}).

Expression (\ref{s2a}) for $s^{(a)}_2$  is found as follows. Since
in the intermediate state 2 in the graph \ref{3b}~(a)
there are three particles (with the
momenta $k_1,k_3,k_4$), according to Eq.~(\ref{sint}) we have:
$$s^{(a)}_2=(k_1+k_3+k_4)^2= \frac{\vec{R}_{1\perp}^2+m^2}{x_1}+
\frac{\vec{R}_{3\perp}^2}{x_3}+
\frac{\vec{R}_{4\perp}^2+m^2}{x_4}, $$ where
$\vec{R}_{1\perp},\vec{R}_{3\perp},\vec{R}_{4\perp}$ are the
perpendicular to $\vec{\omega}$ components of the four-vectors $$
R_1=k_1-x_1p,\quad R_3=k_3-x_3p,\quad R_4=k_4-x_4p.$$ Using the
momenta conservation in the vertices, we transform $R_3$ as:
 $$R_3=k_3-x_3p=k'_1-k_1-(x'_1-x_1)p\;\;(mod\;\;\omega)=
 R'_1-R_1+x'_1Q,$$
where $(mod\;\;\omega)$ means that we omit
the terms, proportional to $\omega$ since they do not contribute
to the $\perp$-components. The variable  $R'_1$ is defined as $R'_1=k'_1-x'_1p'$.
 Hence we get:
$\vec{R}_{3\perp}=\vec{R'}_{1\perp}-\vec{R}_{1\perp}
+x'_1\vec{Q}_{\perp}$, as appears in (\ref{s2a}). Similarly we
find: $\vec{R}_{4\perp}=-\vec{R'}_{1\perp}-x'_1\vec{Q}_{\perp}$
and reproduce in this way Eq.~(\ref{s2a}) for $s^{(a)}_2$.

To find  $s^{(a)}_3$, we start with: $$ s^{(a)}_3=
(k_1+k_3+k'_2)^2= \frac{\tilde{\vec{R}}_{1\perp}^2+m^2}{x_1}+
\frac{\tilde{\vec{R}}_{3\perp}^2}{x_3}+
\frac{\vec{R'}_{1\perp}^2+m^2}{x'_2},$$ where all $R$'s are
defined relative to $p'$: $$\tilde{R}_1=k_1-x_1p',\quad
\tilde{R}_3=k_3-x_3p',\quad R'_1=k'_1-x'_1p'$$ and then, again
using the conservation laws, we find
$\tilde{\vec{R}}_{1\perp}=\vec{R}_{1\perp}-x_1\vec{Q}_{\perp}$,
$\tilde{\vec{R}}_{3\perp}=\vec{R'}_{1\perp} -
\vec{R}_{1\perp}+x_1\vec{Q}_{\perp}$ and derive Eq.~(\ref{s3a})
for $s^{(a)}_3$.

Similarly we obtain the contribution of the diagram Fig.
\ref{3b}~(b):
\begin{eqnarray}\label{ffb} F_{3b}^{(b)}=\frac{16\pi\alpha
m^2}{(2\pi)^6}&&\int
\frac{\psi(\vec{R}_{1\perp},x_1)\;\psi(\vec{R'}_{1\perp},x'_1)}
{(s^{(b)}_2-M^2)(s^{(b)}_3-M^2)}
\nonumber\\
&&\times\frac{\theta(x_1-x'_1)}{(x_1-x'_1)}
\frac{\d^2R_{1\perp}\d x_1}{2x_1(1-x_1)}
\;\frac{\d^2R'_{1\perp}\d x'_1}{2x'_1(1-x'_1)},
\end{eqnarray}
where
\begin{eqnarray}
s^{(b)}_2=
\frac{(\vec{R'}_{1\perp}+x'_1\vec{Q}_{\perp})^2+m^2}{x'_1}+
\frac{(\vec{R}_{1\perp}-\vec{R'}_{1\perp}
-x'_1\vec{Q}_{\perp})^2}{x_1-x'_1}+
\frac{\vec{R}_{1\perp}^2+m^2}{1-x_1},\label{s2b}\\ s^{(b)}_3=
\frac{\vec{R'}_{1\perp}^2+m^2}{x'_1}+
\frac{(\vec{R}_{1\perp}-\vec{R'}_{1\perp}
-x_1\vec{Q}_{\perp})^2}{x_1-x'_1}+
\frac{(\vec{R}_{1\perp}-x_1\vec{Q}_{\perp})^2+m^2}{1-x_1}.\label{s3b}
\end{eqnarray}
By the replacement of variables $\vec{R}_{1\perp}\to\vec{R'}_{1\perp},
x_1\to x'_1$, $\vec{R'}_{1\perp}\to\vec{R}_{1\perp}, x'_1\to x_1$
we show that $F_{3b}^{(b)}=F_{3b}^{(a)}$. Therefore we finally get
for full three-body contribution:
$$F_{3b}(Q^2)=F^{(a)}_{3b}(Q^2)+F^{(b)}_{3b}(Q^2)=2F^{(a)}_{3b}(Q^2)$$
and derive in this way Eq.~(\ref{ffa}) for $F_{3b}(Q^2)$.

\subsubsection{Normalization integral}\label{norm3}
Three-body contribution to the normalization integral
$N_3=F_{3b}(0)$ is given by Eq.~(\ref{N3}). It is instructive to
represent it through the interaction kernel. For this aim, we
rewrite (\ref{N3}) as
 \begin{eqnarray}\label{full} N_3&=& F_3^{(a)}(0)+ F^{(b)}_3(0)
 \nonumber\\
 &=&\frac{\alpha m^2}{4\pi^5}\int \psi(\vec{R'}_{1\perp},x'_1)\;\left[
\frac{\theta(x'_1-x_1)}{(x'_1-x_1)(s_a-M^2)^2}+
\frac{\theta(x_1-x'_1)}{(x_1-x'_1)(s_b-M^2)^2}\right]\;
 \nonumber\\ &&
\times\psi(\vec{R}_{1\perp},x_1)
\frac{\d^2R_{1\perp}\d x_1}{2x_1(1-x_1)}
\;\frac{\d^2R'_{1\perp}\d x'_1}{2x'_1(1-x'_1)} \end{eqnarray} with
\begin{equation}\label{sb}
s_b =\frac{\vec{R'}_{1\perp}^2+m^2}{x'_1}+
\frac{(\vec{R}_{1\perp}-\vec{R'}_{1\perp})^2}{x_1-x'_1}+
\frac{\vec{R}_{1\perp}^2+m^2}{1-x_1} .
\end{equation}

\begin{figure}[!ht]
\centering
 \includegraphics{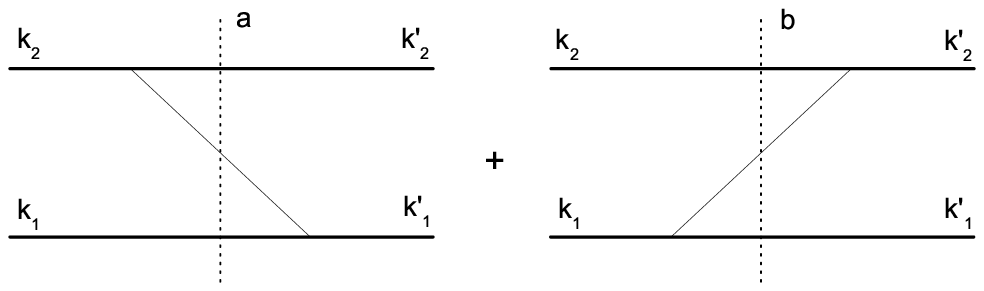}
\centerline{\small Fig.~16. Ladder kernel.}
\end{figure}

The ladder kernel is shown graphically in  Fig.~16. Its
calculation by the Weinberg rules  \cite{sw} gives:
\begin{equation}\label{V}
V=-\frac{4\pi\alpha\;\theta(x'_1-x_1)}{(x'_1-x_1)(s_a-M^2)}
-\frac{4\pi\alpha\;\theta(x_1-x'_1)}{(x_1-x'_1)(s_b-M^2)},
\end{equation}
with $s_a,s_b$ given by Eqs.~(\ref{sa}), (\ref{sb}). The kernel
(\ref{V}) (which enters the Schr\"odinger-type equation) is
related to the amplitude ${\mathcal K}$ of graph Fig.~16 as
$V=-{\mathcal K}/(4m^2)$.

Let us calculate the derivative $\partial V/\partial M^2$:
\begin{equation}\label{dV}
\frac{\partial V}{\partial
M^2}=-4\pi\alpha\;\left[\frac{\theta(x'_1-x_1)}{(x'_1-x_1)(s_a-M^2)^2}
+\frac{\theta(x_1-x'_1)}{(x_1-x'_1)(s_b-M^2)^2}\right].
\end{equation}
Comparing (\ref{dV}) with the integrand of (\ref{full}),
we represent $N_3$, Eq.~(\ref{full}), in the form
 \begin{eqnarray}\label{NN3}
N_3= -\frac{4m^2}{(2\pi)^6}\int
 \psi(\vec{R'}_{1\perp},x'_1)&& \frac{\partial V}{\partial
M^2} \; \psi(\vec{R}_{1\perp},x_1)
\nonumber\\
&&\times\frac{\d^2R_{1\perp}\d x_1}{2x_1(1-x_1)}
\;\frac{\d^2R'_{1\perp}\d x'_1}{2x'_1(1-x'_1)}.
 \end{eqnarray}
Then we make replacement of variables $R_{1\perp},x_1\to
k,\theta$:
\begin{equation}\label{var}
R_{1\perp}=k\sin\theta,\quad
 x_1=\frac{1}{2}\left(1-\frac{k\cos\theta}{\varepsilon_k}\right),
\end{equation}
where $\varepsilon_k=\sqrt{k^2+m^2}$, and similarly for
$R'_{1\perp},x'_1\to k',\theta'$. The integration volume is
transformed as: $$ \frac{\d^2R_{1\perp}\d x_1}{2x_1(1-x_1)}
=\frac{R_{1\perp}\d R_{1\perp}\d x_1d\phi}{2x_1(1-x_1)}=
\frac{k^2dk\sin\theta d\theta d\phi}{\varepsilon_k}=
\frac{\d^3k}{\varepsilon_k}. $$ In these variables $N_3$ obtains
the form:
 \begin{equation}\label{NN3a} N_3= -\frac{4m^2}{(2\pi)^6}\int
 \psi(\vec{k'},\vec{n})\; \frac{\partial V(\vec{k'},\vec{k},\vec{n},M^2)}
{\partial M^2} \;\psi(\vec{k},\vec{n})\;
\frac{\d^3k}{\varepsilon_k}\; \frac{\d^3k'}{\varepsilon_{k'}},
 \end{equation}
where we introduced the vector $\vec{n}$:
$\vec{k}\cd\vec{n}=k\cos\theta$,
$\vec{k'}\cd\vec{n}=k'\cos\theta'$. The equation (\ref{NN3a})
coincides with the many-body contribution (for a general kernal
$V$, not only for the ladder one), given by Eq.~(3.50) in
\cite{cdkm}. It can be obtained by more general method
\cite{cdkm}, without precising the kernel $V$.

For small $\alpha$ the value $N_3$ can be calculated explicitly.
We rewrite $N_3$, Eq.~(\ref{N3}), as:
 \begin{eqnarray}\label{N31}
N_3=\frac{\alpha m^2}{2\pi^5}&&\int
\frac{\psi(\vec{R}_{1\perp},x_1)\;\psi(\vec{R'}_{1\perp},x'_1)(x'_1-x_1)}
{[(s_a-M^2)(x'_1-x_1)]^2}
\nonumber\\
&&\times\theta(x'_1-x_1)\;
\frac{\d^2R_{1\perp}\d x_1}{2x_1(1-x_1)}
\;\frac{\d^2R'_{1\perp}\d x'_1}{2x'_1(1-x'_1)}
\end{eqnarray}
with  $s_a$ represented more explicitly:
\begin{equation}\label{sa1}
s_a= \frac{\vec{R}_{1\perp}^2+m^2}{x_1}+
\frac{{R'}^2_{1\perp}-2R_{1\perp}R'_{1\perp} \cos\phi +
{R}^2_{1\perp}}{x'_1-x_1}+ \frac{\vec{R'}_{1\perp}^2+m^2}{1-x'_1}.
\end{equation}
$\phi$ is angle between $\vec{R}_{1\perp}$ and
$\vec{R'}_{1\perp}$. We make replacement of variables by Eqs.
(\ref{var}). Since $N_3$, Eq.~(\ref{N31}), is proportional to
$\alpha$, for the wave function we can take Eq.~(\ref{psi2m}) for
asymptotically small $\alpha$, where higher degrees of $\alpha$
are neglected. In the variables (\ref{var}) it obtains the form
(cf. Eq.~(3.64) from \cite{cdkm}):
\begin{equation}\label{psir}
\psi(k,\theta)=\frac{8\sqrt{\pi m}\kappa^{5/2}}{(k^2+\kappa^2)^2
\left(1+\frac{k\left|\cos\theta\right|}{\sqrt{k^2+m^2}}\right)},
\end{equation}
where $\kappa=\sqrt{m B}=m\alpha/2$ ($B$ is the binding energy).

Then we make another replacement of variables $k,k'\to p,p'$:
\begin{equation}\label{kp}
 k=\kappa p,\quad k'=\kappa p',
\end{equation}
where $p,p'$ are dimensionless. In the limit $\kappa\to 0$, we
decompose the factor $(s_a-M^2)(x'_1-x_1)$ in the denominator of
(\ref{N31})  in series of $\kappa$, keeping the leading term:
\begin{eqnarray}\label{sa11}
(s_a-M^2)(x'_1-x_1)&\approx&\kappa^2\left[p^2-2p p'
\cos\theta\cos\theta'-2p p' \sin\theta\sin\theta'\cos\phi +{p'}^2
\right. \nonumber\\
&+&\left.\frac{\kappa}{m}(p^2+{p'}^2+2)(p\cos\theta-p'\cos\theta')\right]
\end{eqnarray}
Introducing vector $\vec{q}=\vec{p'}-\vec{p}$, we represent
(\ref{sa11}) as:
\begin{equation}\label{sa2}
(s_a-M^2)(x'_1-x_1)\approx \kappa^2(q^2
-\frac{\kappa}{m}(p^2+{p'}^2+2)\vec{q}\cd\vec{n}) .
\end{equation}
The difference $(x'_1-x_1)$ in the numerator of (\ref{N31}) is
transformed as: $$ x'_1-x_1\approx
\frac{\kappa}{2m}(p\cos\theta-p'\cos\theta')= -\frac{\kappa}{2m}
\vec{q}\cd\vec{n}.$$

Wave function (\ref{psir}) obtains the form:
\begin{equation}\label{psin}
\psi(p)\approx\frac{8\sqrt{\pi m}\kappa^{5/2}}{\kappa^4(p^2+1)^2} .
\end{equation}
It does not depend on $\theta$, this simplifies the calculation.
The integration volume reads:
$$
\frac{\d^3k}{\sqrt{{k}^2+m^2}}\approx\frac{\kappa^3\d^3p}{m}.
$$
Hence, $N_3$ is transformed as:
\begin{equation}\label{N3b}
N_3=-\frac{32\kappa m}{\pi ^4} \int \frac{(\vec{n}\cd\vec{q})\;
\theta(-\vec{n}\cd\vec{q})\; p^2dpdo_{\vec{p}}\;{p'}^2 dp'
do_{\vec{p'}}}
{\left(mq^2-\kappa(p^2+{p'}^2+2)\;\vec{n}\cd\vec{q}\right)^2
(p^2+1)^2({p'}^2+1)^2} .
\end{equation}
One can substitute here denominator in the form of (\ref{sa11}) and
integrate over three angles $\theta,\theta'$ and $\phi$. However, the same
result can be found in a more simple way. Namely, since $N_3$ in
(\ref{N3b}) does not depend on $\vec{n}$, but the integrand depends on
$\vec{n}$, one can at first average integrand over the
$\vec{n}$-directions. Therefore we replace the factor:
$$
h_1=\frac{(\vec{n}\cd\vec{q})\; \theta(-\vec{n}\cd\vec{q})}
{\left(mq^2-\kappa(p^2+{p'}^2+2)\;\vec{n}\cd\vec{q}\right)^2 }
$$
by
$$
h_2=\int \frac{(\vec{n}\cd\vec{q})\; \theta(-\vec{n}\cd\vec{q})}
{\left(mq^2-\kappa(p^2+{p'}^2+2)\;\vec{n}\cd\vec{q}\right)^2 }
\;\frac{do_{\vec{n}}}{4\pi}.
$$
The angle integration here is
one-dimensional and $h_2$ is easily calculated. After that the angle
integrations in (\ref{N3b}) are also carried out easily. Then the
integrals over $p$ and $p'$ are calculated approximately, using that
$\alpha\ll 1$ implies $\kappa\ll m$. In this way we obtain the analytical
formula (\ref{N3s}) for $N_3$ at $\alpha\ll 1$.


\end{document}